\documentclass[aps,prb,twocolumn,10pt,superscriptaddress,notitlepage,floatfix]{revtex4-1}
\usepackage{graphicx}
\usepackage[caption=false]{subfig}
\usepackage[normalem]{ulem}
\usepackage{float}
\usepackage{natbib}
\usepackage{xcolor}
\usepackage{amsmath}
\usepackage{amssymb}
\usepackage{array}
\usepackage{wasysym}
\usepackage{color,soul}
\usepackage{braket}
\usepackage{verbatim}
\usepackage{hyperref}

\graphicspath{{./figures/}}

\begin{document}

\title{Magnetic order and single-ion anisotropy in Tb$_3$Ga$_5$O$_{12}$}

\author{R.~Wawrzy\'{n}czak}
\email{wawrzynczak@ill.fr}
\affiliation{Institut Laue-Langevin, CS 20156, Cedex 9, 38042 Grenoble, France}
\affiliation{Laboratory for Neutron Scattering and Imaging, Paul Scherrer Institut, 5232 Villigen PSI, Switzerland}
\author{B.~Tomasello}
\email{brunotomasello83@gmail.com}\email{tomasello@ill.fr}
\affiliation{Institut Laue-Langevin, CS 20156, Cedex 9, 38042 Grenoble, France}
\author{P.~Manuel}
\affiliation{ISIS Facility, Rutherford Appleton Laboratory-STFC, Chilton, Didcot OX11 0QX, United Kingdom}
\author{D.~Khalyavin}
\affiliation{ISIS Facility, Rutherford Appleton Laboratory-STFC, Chilton, Didcot OX11 0QX, United Kingdom}
\author{M.~D.~Le}
\affiliation{ISIS Facility, Rutherford Appleton Laboratory-STFC, Chilton, Didcot OX11 0QX, United Kingdom}
\author{T.~Guidi}
\affiliation{ISIS Facility, Rutherford Appleton Laboratory-STFC, Chilton, Didcot OX11 0QX, United Kingdom}
\author{A.~Cervellino}
\affiliation{Swiss Light Source, Paul Scherrer Institute, 5232 Villigen PSI, Switzerland}
\author{T.~Ziman}
\affiliation{Institut Laue-Langevin, CS 20156, Cedex 9, 38042 Grenoble, France}
\affiliation{LPMMC, UMR-5493, Universit\'{e} Grenoble Alpes and CNRS, 38042 Grenoble, France}
\affiliation{Kavli Institute for Theoretical Sciences, University of Chinese Academy of Sciences, Beijing 100190, China}
\author{M.~Boehm}
\affiliation{Institut Laue-Langevin, CS 20156, Cedex 9, 38042 Grenoble, France}
\author{G.~J.~Nilsen}
\affiliation{ISIS Facility, Rutherford Appleton Laboratory-STFC, Chilton, Didcot OX11 0QX, United Kingdom}
\author{T.~Fennell}
\email{tom.fennell@psi.ch}
\affiliation{Laboratory for Neutron Scattering and Imaging, Paul Scherrer Institut, 5232 Villigen PSI, Switzerland}

\date{\today}

\begin{abstract}
	Terbium gallium garnet (TGG), Tb$_3$Ga$_5$O$_{12}$, is well known for its applications in laser optics, but also exhibits complex low-temperature magnetism that is not yet fully understood. Its low-temperature magnetic order is determined by means of time-of-flight neutron powder diffraction. It is found to be a multiaxial antiferromagnet with magnetic Tb$^{3+}$ ions forming six sublattices of magnetic moments aligned parallel and anti-parallel to the $\langle100\rangle$ crystallographic directions of the cubic unit cell. The structure displays strong easy-axis anisotropy with respect to a two-fold axis of symmetry in the local orthorhombic environment of the Tb$^{3+}$ sites. The crystal-field splitting within the single-ion ground-state manifold is investigated by inelastic neutron scattering on powder samples. A strong temperature dependence of the quasidoublet ground-state is observed and revised parameters of the crystal-field Hamiltonian are given. The results of bulk magnetic susceptibility and magnetisation measurements are in good agreement with values based on the crystal-field model down to 20~K, where the onset of magnetic correlations is observed. 
\end{abstract}

\maketitle

\section{Introduction}\label{intro}
\begin{figure}
\centering
\includegraphics[width=\linewidth]{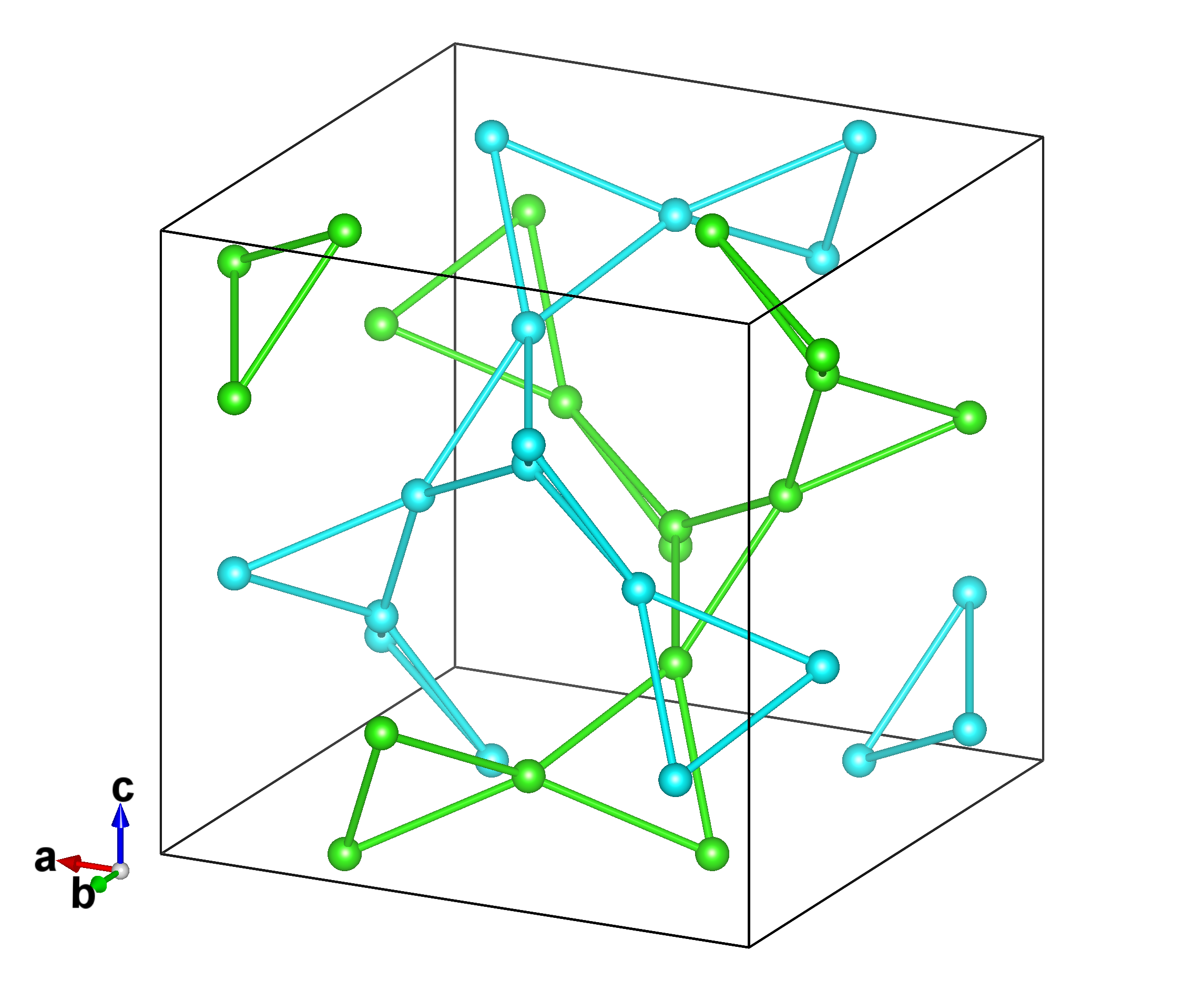}
\caption{The two hyperkagome sublattices of Tb$^{3+}$ ions in TGG. Different colours emphasise that the two inter-penetrating half-garnet lattices are not connected with each other at the nearest neighbour distance. Only links within one unit cell are shown. The top-left and bottom-right triangles (green and cyan respectively) are not isolated, but are linked to their respective sublattices in neighbouring unit cells.~\cite{vesta}.}
 \label{tgg_struc}
\end{figure}

	Many current studies of the garnet family focus on technological applications of Y$_3$Al$_5$O$_{12}$ and Y$_3$Fe$_5$O$_{12}$ (in optoelectronics~\cite{kamal15} and magnonics ~\cite{serga10,collet16,flebus17} respectively), but the garnet structure also contains two inter-penetrating half-garnet lattices (Fig.~\ref{tgg_struc}) - twisting spatial arrangements of corner-sharing triangles - which are of considerable interest for highly frustrated magnetism. An individual half-garnet or hyperkagome lattice, along with the pyrochlore lattice, are candidates for the construction of three dimensional Coulomb phases~\cite{henley10}. There are two series of rare earth garnets, $R_3$Al$_5$O$_{12}$ and $R_3$Ga$_5$O$_{12}$ (where $R$ is a trivalent rare earth ion), and, as in rare earth pyrochlores~\cite{gardner10}, competition between exchange and dipolar interactions\cite{ball63,felsteiner81} and crystal field anisotropy produces contrasting properties throughout the series, often with highly anisotropic magnetic order at low temperature~\cite{ball61,hastings65,hammann68a,hammann68b,hammann69,hammann75,hammann77}. Gd$_3$Ga$_5$O$_{12}$ (GGG)~\cite{hov80} is the only fully frustrated garnet in which no long-range magnetic order has been observed to the lowest temperature~\cite{hastings65,hammann69,Quilliam2013}, and recent investigations discovered cooperative magnetic multipolar degrees of freedom underlying the spin correlations of the low temperature phase of GGG~\cite{paddison15}. 

	In TGG, the non-Kramers ion Tb$^{3+}$, has a quasidoublet ground state composed of two closely spaced singlets as a result of the orthorhombic crystal electric field (CEF)\cite{guillot85}. Thus the magnetic properties of TGG are strongly anisotropic, and long range order is expected to be of the induced-moment type. Highly anisotropic magnetic properties, including step-like magnetisation curves, demonstrate the importance of single-ion crystal field effects~\cite{guillot85,kamazawa08,araki08}. Rare earth garnets with non-collinear Ising anisotropy and dominant dipolar interactions are expected to order with a magnetic structure formed of interpenetrating antiferromagnetic chains running along the $\langle100\rangle$ directions~\cite{ball63,hastings65,hammann69}. TGG orders with this structure at $T_{\rm N}\approx0.25$~K~\cite{hammann73}, but it has been suggested that hyperfine interactions must be incorporated to explain the observed $T_{\rm N}$~\cite{hammann73,hammann75,hammann77}. The existence of a correlated phase for $T>T_{\rm N}$ has not been examined. 

	Moreover, at higher temperatures, in addition to its well-known large Verdet constant and transparency, which make it a suitable material for Faraday rotators and optical isolators\cite{dentz74,villaverde78,barnes92}, TGG also exhibits an acoustic Faraday effect~\cite{thalmeier09,sytcheva10} Cotton-Mouton~\cite{Low2014} and a thermal Hall effect~\cite{strohm05,inyushkin07} (for which it is the prototypical system), suggestive of important spin-lattice interactions. Despite several attempts~\cite{sheng06,kagan08,mori14}, the thermal Hall effect lacks an unequivocal microscopic description. Detailed knowledge of the magnetic interactions and crystal field wavefunctions in TGG plays a crucial role as a starting point required to understand any frustrated phase and the ordering mechanism that terminates it, as well as an eventual understanding of the magnetoelastic and magnetothermal effects. 

	Here, results of our studies on powder samples of TGG are presented. Neutron and x-ray diffraction were used to determine details of the crystalline and magnetic structures (Sec. \ref{sec:struc}) that are fundamental for any further considerations. Inelastic neutron scattering (INS) was employed to observe crystal electric field transitions (Sec.\ref{sec:cef}) and this information was used to refine the parameters of the CEF Hamiltonian. The results of bulk magnetisation measurements are compared to the calculated forms determined by the CEF model (Sec. \ref{sec:mag}). Discussion of the results is followed by remarks as to the outlook for future single crystal studies. 
		
\section{Methods}\label{methods}
\subsection{Sample preparation} 
	A single crystal in the form of a cylindrical rod (\diameter =$6$~mm and $l\sim{}50$~mm) cut out of an ingot grown by the Czochralski method was acquired from FEE (Idar-Oberstein, Germany) and crushed to a fine powder in an agate pestle and mortar.

\subsection{Synchrotron x-ray powder diffraction} 
	For a precise determination of structural parameters, a high-resolution synchroton x-ray powder diffraction (SXRPD) experiment was performed on the Materials Science Beamline at the Swiss Light Source (Paul Scherrer Institute, Villigen, Switzerland). The powder sample was loaded in a borosilicate glass capillary and a diffraction pattern was taken at room temperature ($298$~K). The \diameter =$0.1$~mm diameter of the capillary in combination with photon energy $E=22$ keV ($\lambda=0.563$~\AA{}) were chosen to minimise the sample absorption.

\subsection{Neutron time-of-flight powder diffraction} 
	Neutron diffraction patterns were collected on the WISH time-of-flight diffractometer at ISIS neutron and muon facility (Rutherford-Appleton Laboratory, Harwell, UK). A dilution refrigerator insert was used to cool the sample below $T_{\rm N}$. For this procedure $\sim{}5.5$~g of powder was compacted in a $\diameter{}=5$~mm copper can. In order to enable faster thermalisation of the sample, the can was indium sealed and equipped with a capillary which allowed for the introduction of $^4$He exchange gas during the cooling process. This is particularly important given the low thermal conductivity of garnets at $T<1$~K~\cite{inyushkin10}, the slow thermalisation of their magnetic structures, and the low $T_{\rm N}$ of TGG. The sample was held at base temperature, $T\approx0.03$~K for $\sim{}12$ hours before diffraction measurements commenced. During initial runs, the solid methane moderator was above its intended operating temperature, which modifies the wavelength spectrum and reduces the neutron flux of the instrument, but subsequently diffraction patterns were recorded at base temperature ($T_{base}\sim{}0.03$~K), $1.5$, $5$, and $15$~K. The appearance of magnetic Bragg reflections indicated the thermalisation of at least part of the sample below $T_{\rm N}$, but a significant decrease in the intensity of the magnetic peaks with concurrent development of diffuse scattering around the Bragg positions, was observed during the collection of the base temperature diffraction patterns once the moderator reached full performance. Since the changes occurred only concurrently with the exposure of the sample to the neutron beam, they were attributed to beam heating. This implies that the measured low temperature data are an average over a range of temperatures, and the full low temperature ordered moment cannot be accurately determined. The collected patterns were analysed by the Rietveld and the LeBail refinement methods with use of the \texttt{FULLPROF} suite~\cite{rodriguez93}.

\subsection{Inelastic neutron scattering} 
	Time-of-flight neutron spectroscopy experiments were conducted on the MARI (direct geometry) and IRIS (indirect geometry) time-of-flight spectrometers at ISIS (Rutherford-Appleton Laboratory, Harwell, UK). For MARI, $7.3$~g of powder was wrapped in an aluminium foil sachet and placed in a $\diameter{}=45$~mm Al can. Spectra with $E_i=150,75,50$ and $12$~meV were recorded at $T=5$~K and $T=100$~K, using a Fermi chopper with a gadolinium coated slit package. The best energy resolution at energy transfer $E\sim{}5$~meV was obtained with $E_i = 12$~meV, where $\Delta{}E\approx0.19$~meV. For IRIS, the same amount of sample, also wrapped in an Al sachet, was placed in a $\diameter{}=24$~mm can. Datasets were collected with the PG(002) and PG(004) analyser setups. The former allows for measurement within an almost symmetric, narrow dynamical range ($-0.55<E<0.57$~meV) and its excellent resolution of $17.5$~$\mu$eV at the elastic line was used to investigate the components of the ground-state quasidoublet. The latter setup expands the dynamical window on the neutron energy loss side ($-2.1<E<15.5$~meV), providing a $54.5$~$\mu$eV resolution at the elastic line, thus enabling observations of CEF excitations at higher energies. The measurements were conducted at temperatures in range $1.8<T<100$~K. 

\subsection{Magnetization} 
	Magnetic susceptibility data were collected on a small portion of the same powder sample ($m=95.6$ mg), in an applied magnetic field of $0.01$~T, using a Quantum Design MPMS XL-7 superconducting quantum interference device (SQUID) magnetometer. Using the same device, the magnetisation was measured with a smaller sample ($m=8.15$~mg) fo fields up to $6$~T and temperature range $1.8<T<300$~K. 

\section{Results}
\subsection{Crystalline and ordered magnetic structure}\label{sec:struc}
\begin{figure}
\centering
\includegraphics[width=\linewidth]{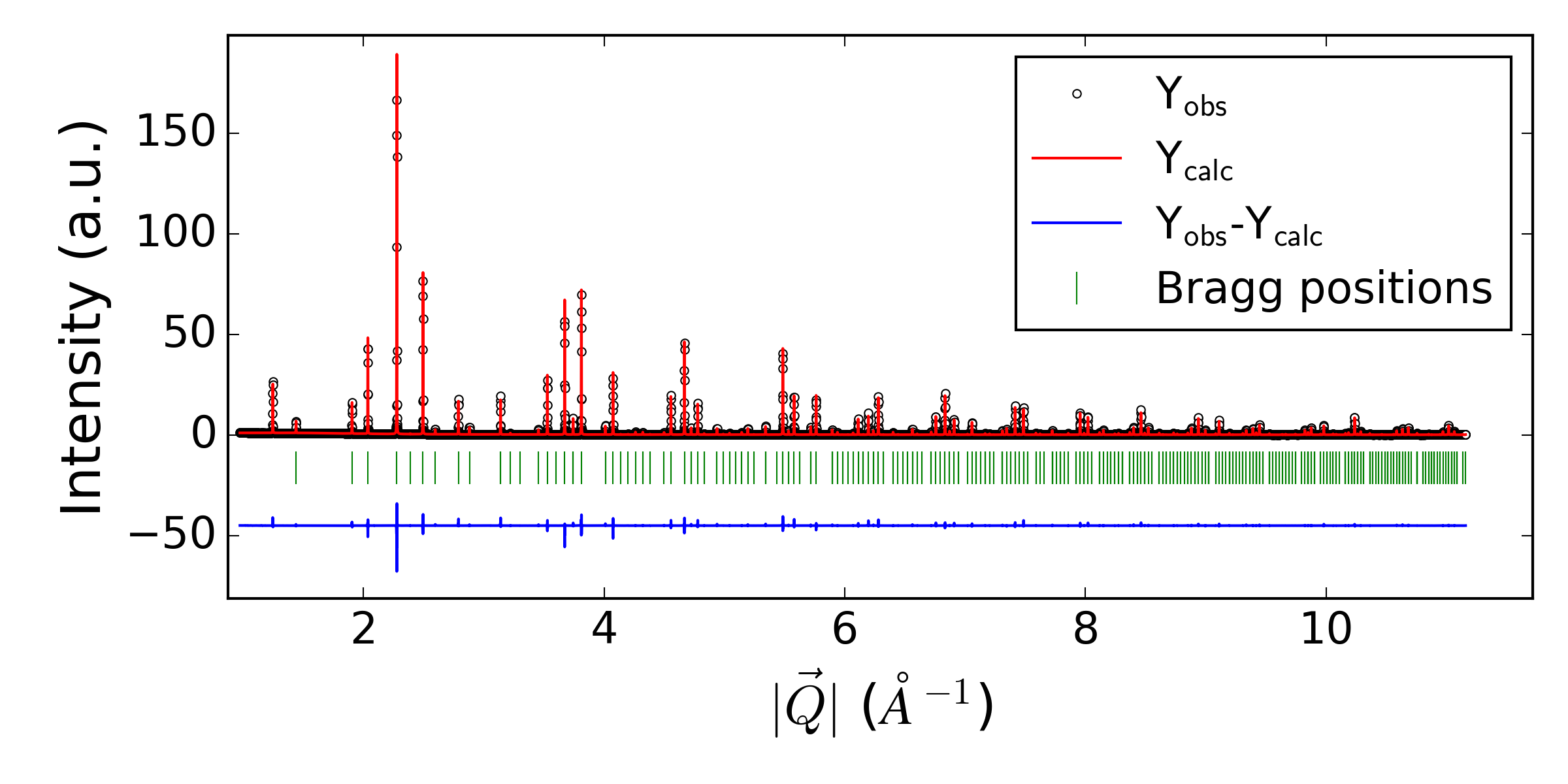}
\caption{Rietveld refinement of synchrotron x-ray powder diffraction pattern. Green vertical lines mark the positions of Bragg reflections.}
\label{xray_pattern}
\end{figure}

	TGG crystallises in the body-centred cubic $Ia\bar{3}d$ space-group (No. 230). The Tb$^{3+}$ ions occupy the 24$c$ Wyckoff positions (with point group $D_2$ or $222$), forming two half-garnet sub-lattices (Fig.~\ref{tgg_struc}). The lattice constant of TGG at room temperature, determined from a Rietveld refinement of the x-ray diffraction pattern is $a=12.35225(1)$~\AA . The presence of superstoichiometric Tb$^{3+}$ ions in Czochralski-grown TGG has been discussed in the literature~\cite{mori14}. Because of the large contrast between Ga$^{3+}$ and Tb$^{3+}$ in x-ray diffraction, Rietveld refinement enabled a precise determination of the stoichiometry of our sample, which was found to be Tb$_{3.031(3)}$Ga$_{4.969(3)}$O$_{12}$. (The small difference between the coherent neutron scattering lengths of Tb and Ga does not allow for cross-verification by neutron diffraction~\cite{Note5}). Superstoichiometric Tb$^{3+}$ ions were found exclusively at the octahedral Ga site $16a$ (with trigonal point group $C_{3i}$ or $\bar{3}$). Attempts to refine a population of terbium ions on other sites led to non-physical negative values of the site occupancy. The rest of the refined crystal structure parameters are presented in Table \ref{cryst_param} and the Rietveld refinement is presented in Fig.~\ref{xray_pattern}. 

\begin{table*}
\caption{Structural parameters obtained by Rietveld refinement of SXRPD pattern measured at room temperature. Tb2 stands for superstoichiometric Tb$^{3+}$ ions at the $16a$ gallium position. \label{cryst_param}}
\begin{ruledtabular}
\begin{tabular}[c]{ccccccc}
Atom	& Wyckoff position	& $x$			& $y$			& $z$			& Occupancy	& $B_{iso}$ (\AA$^2$)	\\ 
\hline
Ga1		& $16a$				& 0 			& 0				& 0				& 1.969(3) 	& 0.217(5)				\\
Tb2* 	& $16a$				& 0 			& 0				& 0				& 0.031(3)	& 0.217(5)				\\
Tb		& $24c$				& 1/8			& 0				& 1/4			& 3			& 0.239(1)				\\
Ga2		& $24d$				& 3/8			& 0				& 1/4			& 3			& 0.283(3)	 			\\
O		& $96h$				& 0.286(1)		& 0.0566(1)		& 0.6500(1)		& 12		& 0.23(2)				\\
\end{tabular}
\end{ruledtabular}
\end{table*}
	
\begin{figure}
\centering
\includegraphics[width=\linewidth]{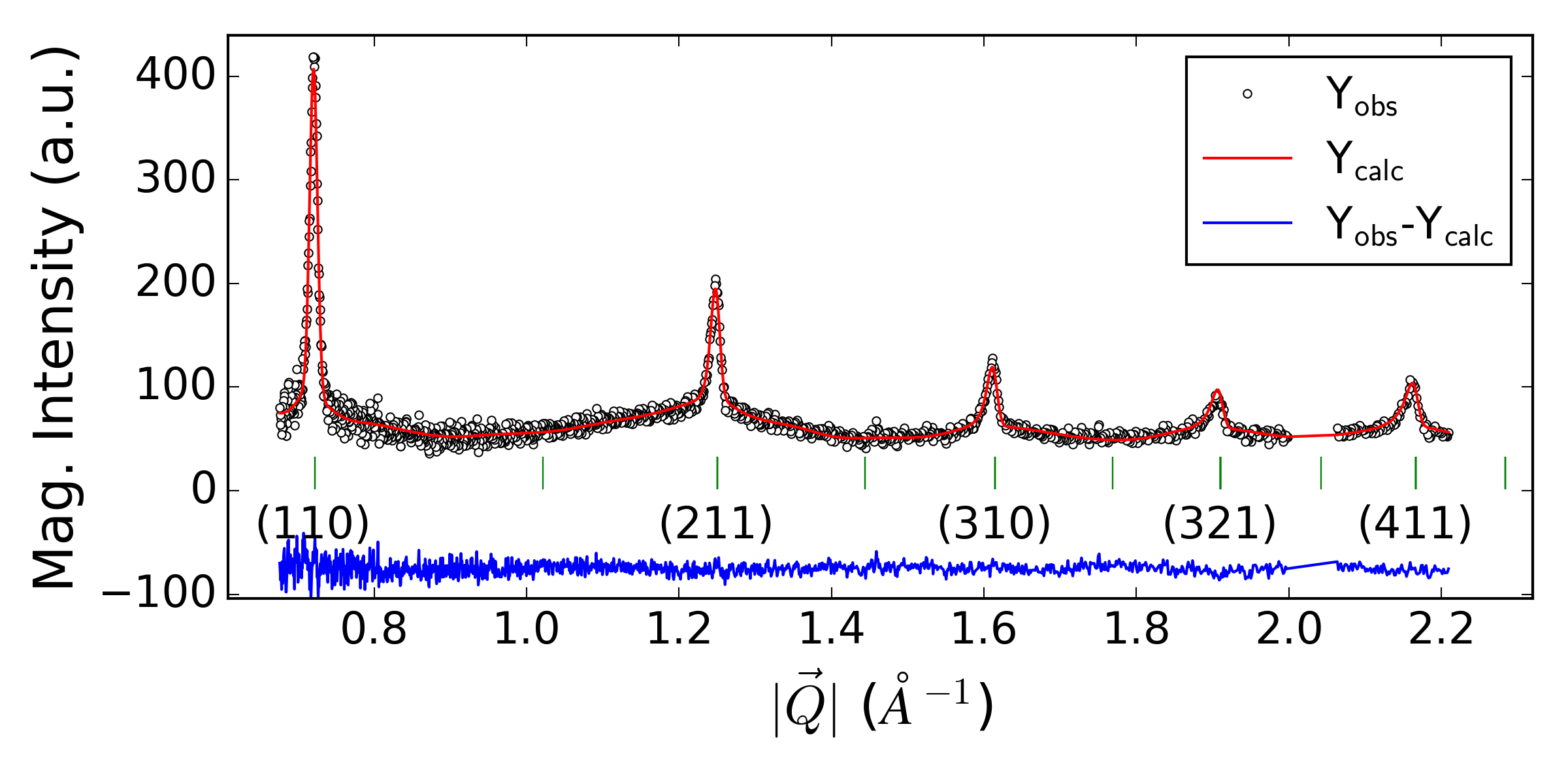}
\caption{Rietveld refinement of magnetic reflections in the neutron powder diffraction difference pattern, produced by the $m\Gamma_{2}^{+}$ irreducible representation of $Ia\bar{3}d$, corresponding to $Ia\bar{3}d'$ magnetic space group. The observed reflections are indexed and can be seen to correspond to the propagation vector $\mathbf{k}=(0,0,0)$. }
\label{mag_int}
\end{figure}

	The magnetic intensity in the neutron diffraction data was separated by subtracting the 15~K dataset from the one measured at $T_{base}$. A flat background was added to remove negative intensity resulting from the presence of diffuse scattering in the vicinity of magnetic Bragg positions at higher temperatures. This procedure allowed for the observation of five magnetic reflections (Fig.~\ref{mag_int}), two of them appearing at nuclear allowed positions $(2,1,1)$ and $(3,2,1)$). All of the magnetic Bragg peaks can be indexed by the propagation vector $\mathbf{k}=(0,0,0)$.
	
	Using the \texttt{ISOTROPY} package~\cite{isotropy}, 8 irreducible representations of the space-group and the propagation vector were found~\cite{kpoint} ($2\times1$-dimensional: $m\Gamma_{2}^{+}$, $m\Gamma_{2}^{-}$; $2\times 2$-dimensional: $m\Gamma_{3}^{+}$, $m\Gamma_{3}^{-}$; $4\times3$-dimensional: $m\Gamma_{4}^{+}$, $m\Gamma_{4}^{-}$, $m\Gamma_{5}^{+}$ and $m\Gamma_{5}^{-}$; in Miller-Love notation). Their magnetic moment basis functions were projected out and combined with corresponding order parameter directions. These generated 14 possible sets of basis vector directions. 11 sets were immediately excluded as they are incompatible with the presence or absence of certain magnetic Bragg peaks in the experimental diffraction pattern, i.e. the diffraction data show the presence of $(1,1,0)$ antiferromagnetic and absence of $(2,0,0)$ ferromagnetic reflections (Fig.~\ref{mag_int}). Rietveld refinement of the remaining possibilities, and examination of the relative intensities of the observed reflections, allowed us to unambiguously assign the 1-dimensional irreducible representation $m\Gamma_{2}^{+}$ to the observed pattern (Fig.~\ref{mag_int}). This structure has the magnetic space-group $Ia\bar{3}d'$. The scale factor of the Rietveld refinements was set by refining the room temperature structural model obtained from the x-ray experiment against the data measured at $15$~K (not shown, no unexpected changes of parameters or misfits occur). In the course of the magnetic intensity refinement, the wavelength-dependent absorption correction parameters were re-refined to provide a better match in the low-$|\vec{Q}|$ region. These adjustments permitted an excellent correspondence of the observed and calculated magnetic intensities in this limited $|\vec{Q}|$-range. The strength of the ordered magnetic moment was determined to be $|\mathbf{m}|=2.16(9)$~$\mu_{\mathrm{B}}$ (where $\mathbf{m} = g_{J} \hat{\mathbf{J}}\mu_{\mathrm{B}}$ with $\hat{\mathbf{J}} \equiv (\hat{J}_{x} , \hat{J}_{y}, \hat{J}_{z})$ being the angular momentum operators and $\mu_{\mathrm{B}}={e\hbar}/{2m_e}$ the Bohr magneton). Such value is significantly smaller than that previously reported for terbium garnets~\cite{hammann73,hammann75} but consistent with the heating of the sample toward $T_{\rm N}$ by the neutron beam (the initial measurement with partially thermalised moderator and inferior refinement gives $|\mathbf{m}|=3.20(9)$~$\mu_{\mathrm{B}}$).
	
\begin{figure}
\centering
\includegraphics[width=0.5\columnwidth]{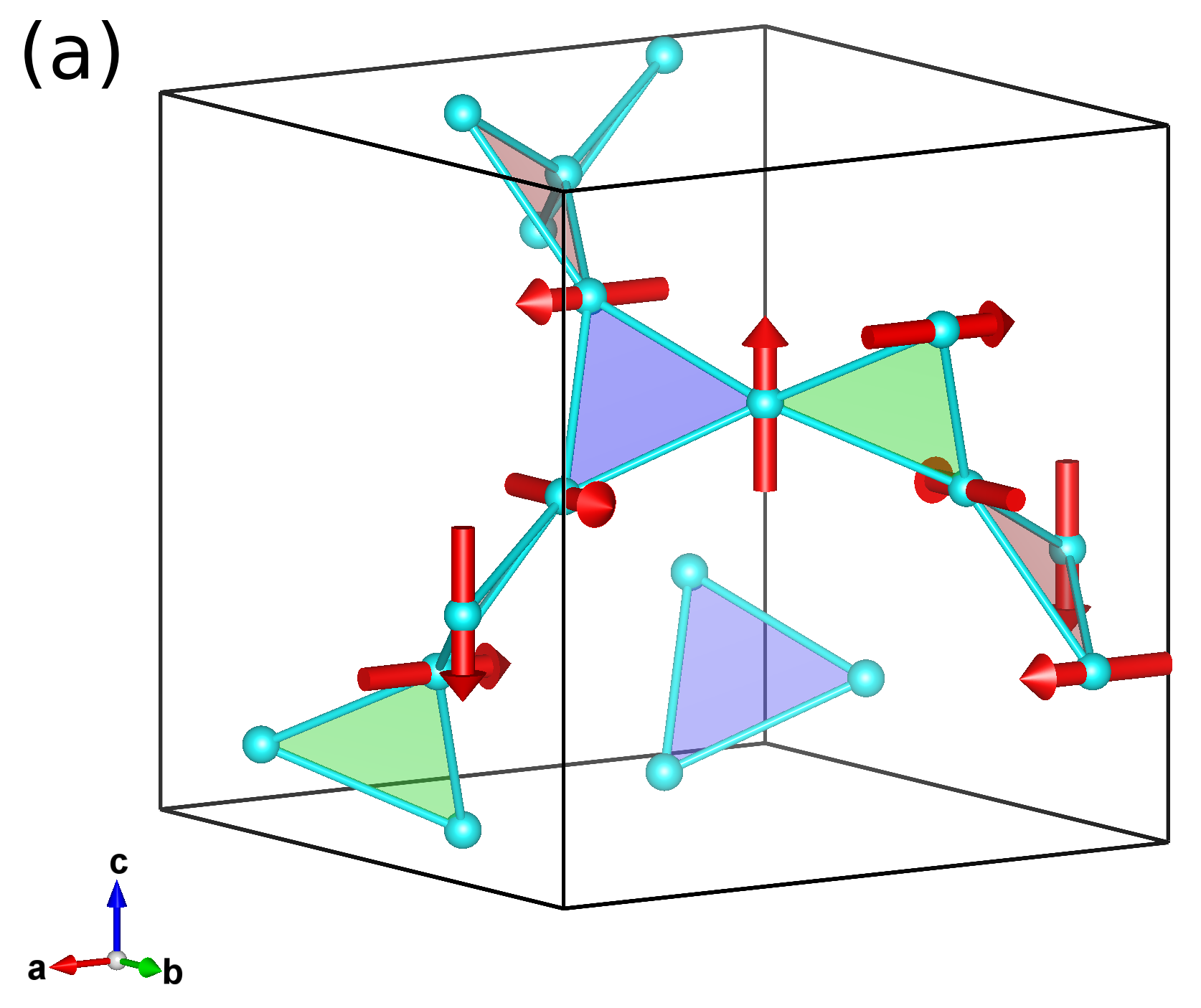}\includegraphics[width=0.5\columnwidth]{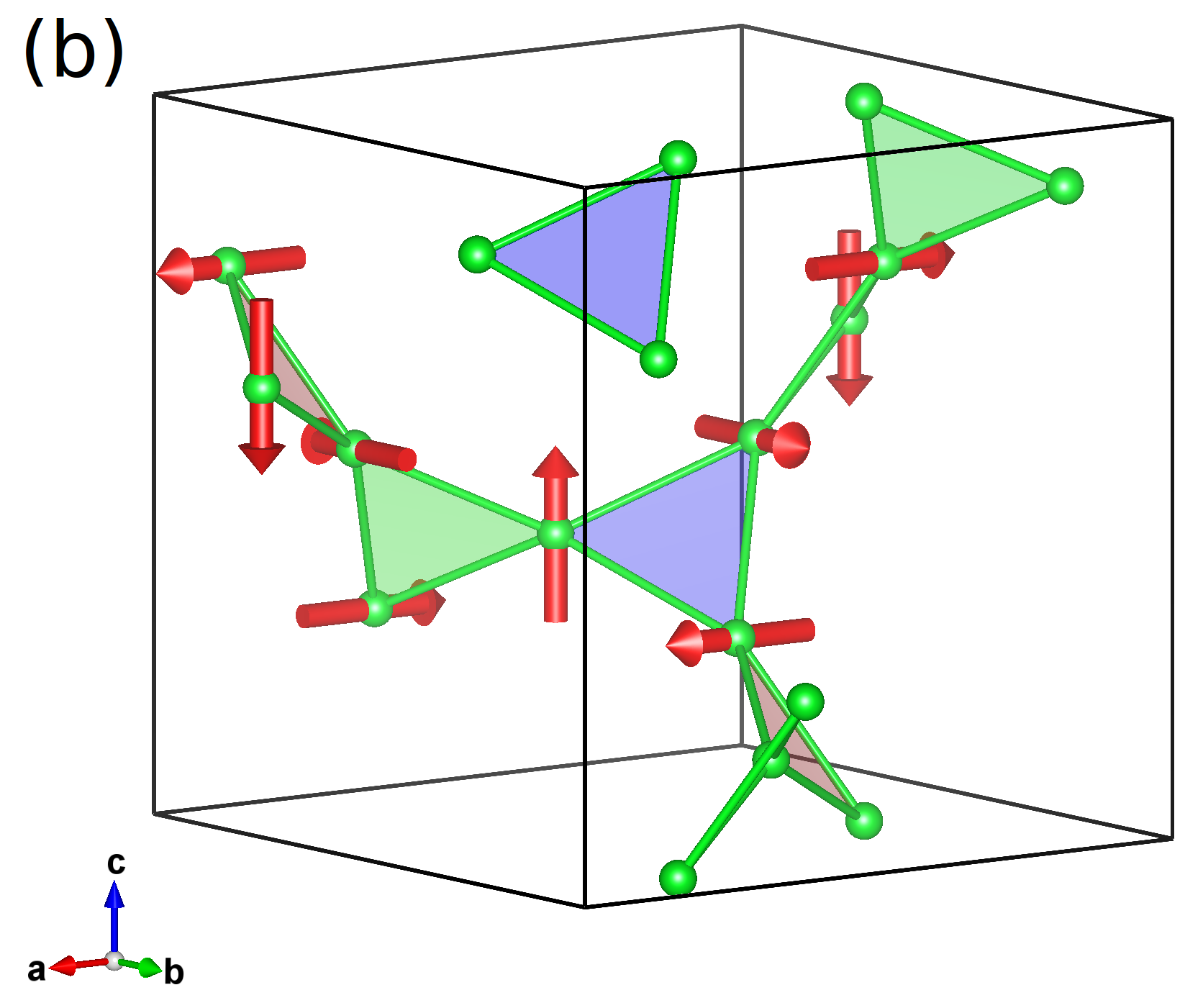}\\\includegraphics[width=0.5\columnwidth]{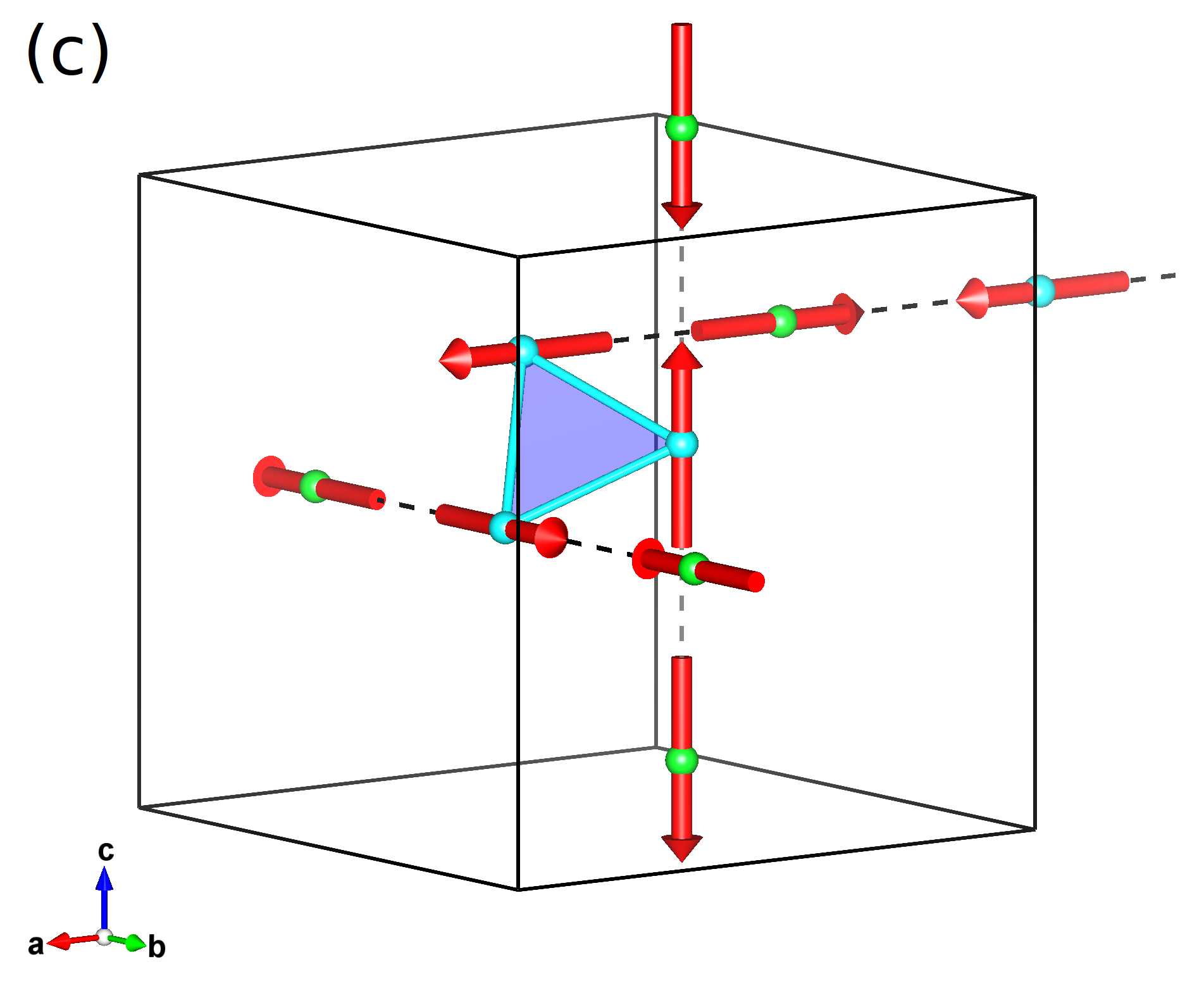}\includegraphics[width=0.5\columnwidth]{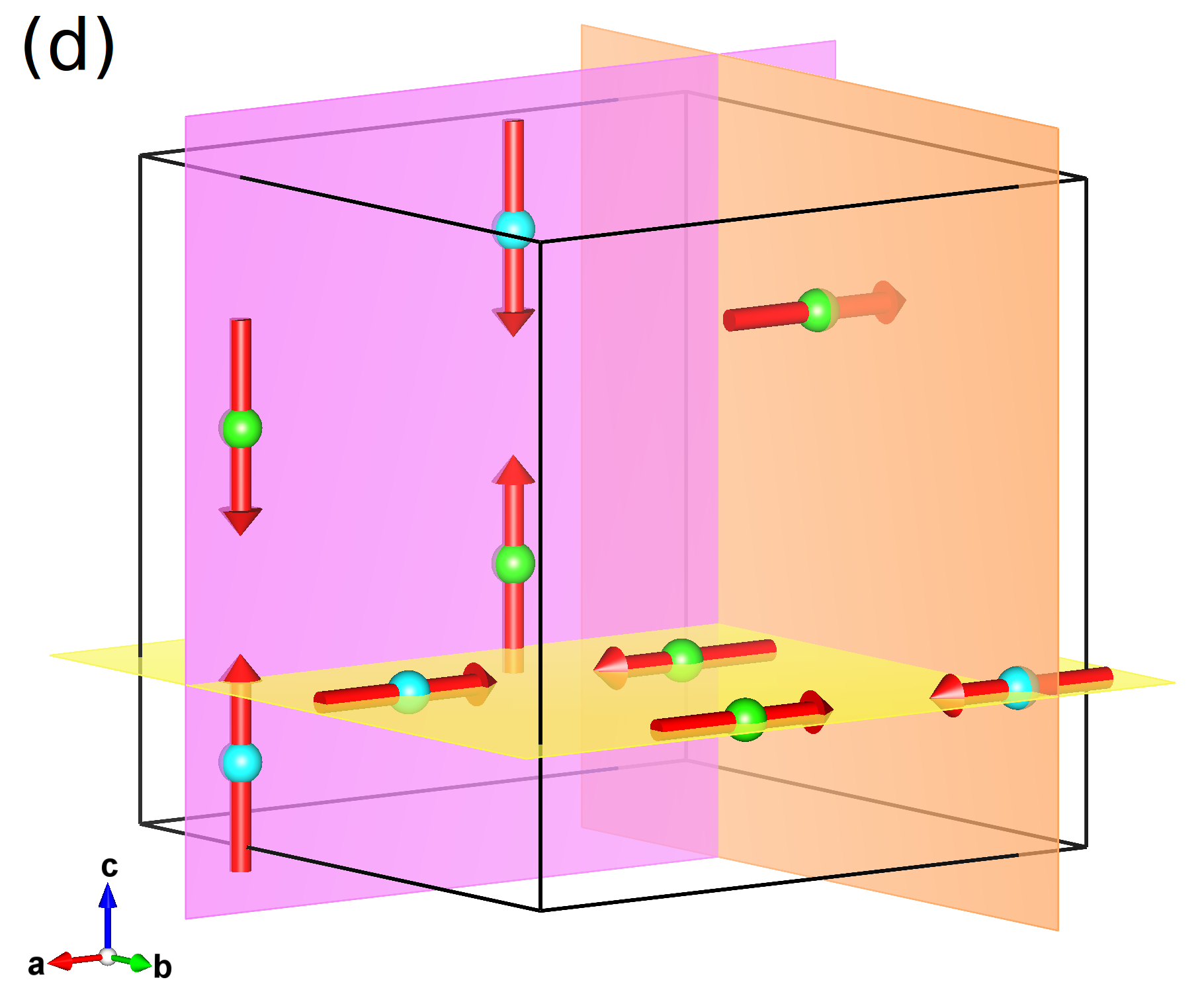}
\caption{Ordered magnetic structure of TGG. (a) and (b), red arrows show the magnetic moments for a selection of the garnet sites from each of the two hyperkagome sublattices. For sites where magnetic moments are not shown, their anisotropy can be inferred by the ones shown on those triangles of the same sublattice that lie in parallel planes. (Such triangles also have the same colour-shading in both panels (a) and (b)). (c) antiferromagnetically ordered chains along principal directions of cubic unit cell (d) basal planes of that cell populated with collinear moments~\cite{vesta}.}
\label{mag_struc}
\end{figure}

	One of the three magnetic moments at the sites forming a single triangle in the half-garnet lattices points along each of the the $\langle100\rangle$ crystal axes, so that they are mutually perpendicular. The selected $\langle100\rangle$ at a particular site corresponds to a 2-fold axis of the $D_2$ point group (the other two of which point along perpendicular $\langle110\rangle$-type directions). The full magnetic structure consists of interpenetrating planes perpendicular to the $\langle100\rangle$ crystal axes, populated with collinear magnetic moments. Co-planar triangles in the half-garnet network connected by the inversion center have identical spin configurations but both types of $\langle\pm100\rangle$ moments appear on each half-garnet lattice, making them individually antiferromagnetic. In total there are six sublattices hosting ions carrying moments pointing along one of the $\langle\pm100\rangle$-type directions. Different visualisations of the structure are presented in Fig.~\ref{mag_struc}.

\subsection{Crystal electric fields in TGG}	\label{sec:cef}
\subsubsection{Inelastic Neutron Scattering Results} \label{sec:local}
\begin{figure}
\centering
\includegraphics[width=\linewidth]{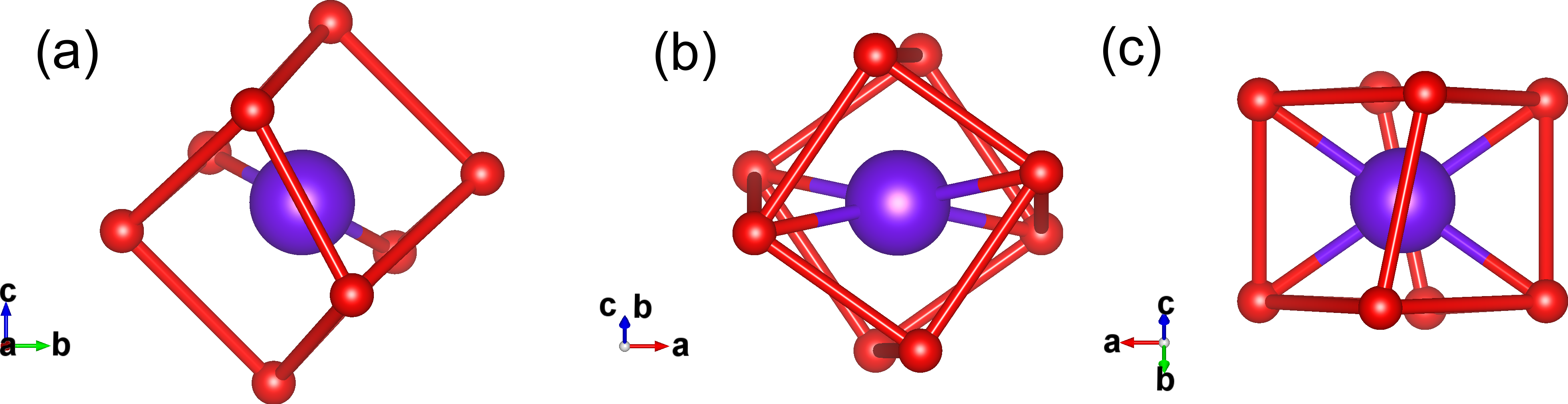}
\caption{Local environment of Tb$^{3+}$ ion (purple spheres) located at ($1/8,0,1/4$) and surrounded by eight O$^{2-}$ ions (red spheres). Projections along three two-fold axis of $D_2$ point group: (a) $\langle100\rangle$, 	(b) $\langle01\bar{1}\rangle$ and (c) $\langle011\rangle$, constituting local coordinates at one of the sites (i.e. $\mathbf{x}_{3}$, $\mathbf{y}_{3}$ and $\mathbf{z}_{3}$ in Eq. \ref{eq:LocCoord3}). The directions of symmetry axes at five other inequivalent sites are reproduced by rotational symmetry elements of $Ia\bar{3}d$. Bicolor cylinders represent shorter of the two Tb-O distances~\cite{vesta}.}
\label{tb_site}
\end{figure}	

	In TGG, each Tb$^{3+}$ ion (at the $24c$ Wyckoff positions) sits at the centre of a distorted cubic cage of eight oxygens, as illustrated in Fig.~\ref{tb_site}. The orthorhombic point-symmetry ($D_2$ or $222$) {predicts} the complete lifting of the degeneracy, i.e. the ground-state $^7F_6$ ($^{2S+1}L_{J}$) term splits into 13 singlets with the symmetry decomposition $4\Gamma_1+3\Gamma_2+3\Gamma_3+3\Gamma_4$ ($\Gamma_{\alpha}$, with $\alpha=1,2,3,4$, are the irreducible representations of the $D_2$ point-group). In our inelastic neutron scattering data we can identify eleven of the twelve transitions between the ground-state and excited states.

\begin{figure}
\centering
\includegraphics[width=\linewidth]{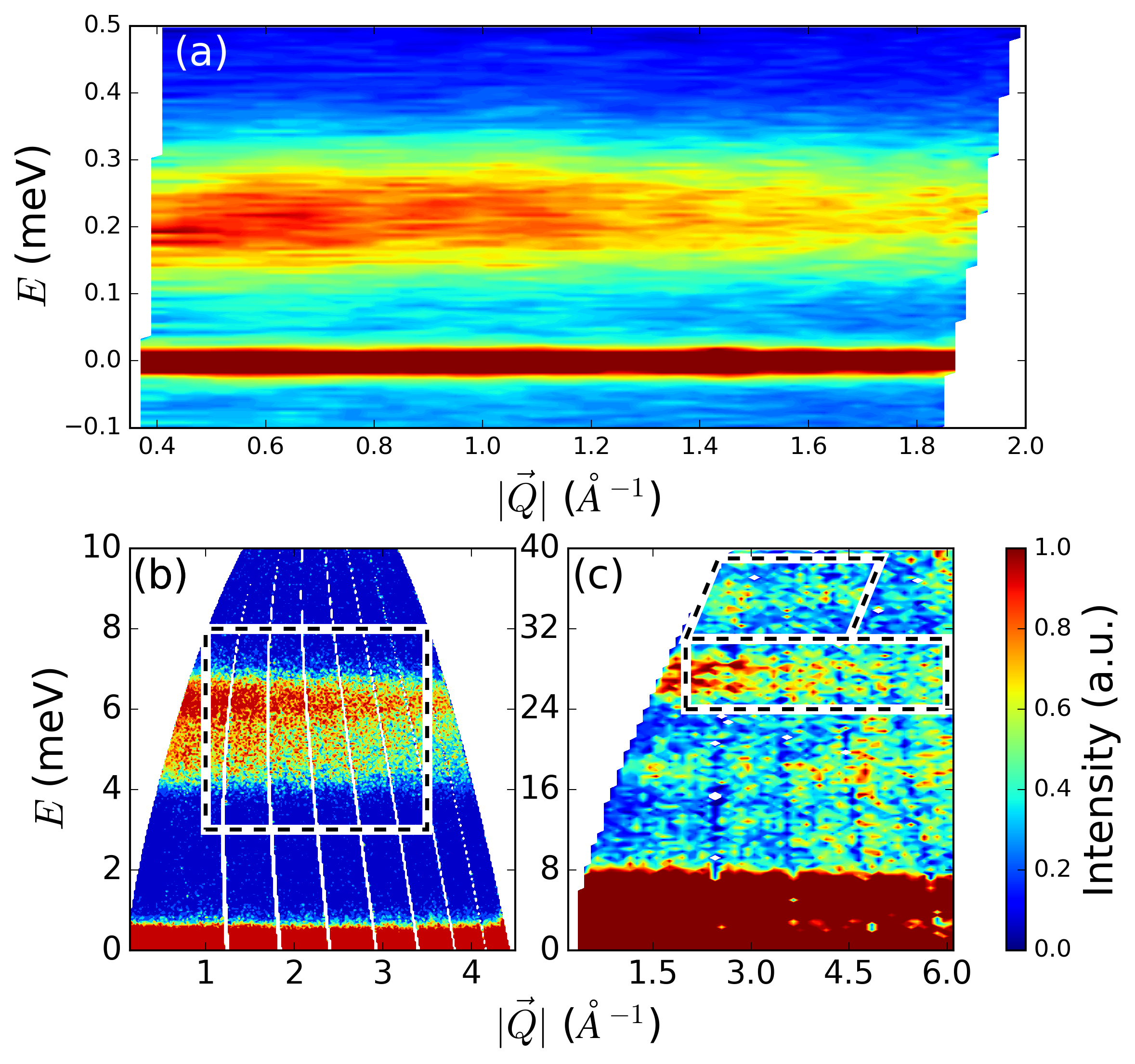}
\caption{Powder neutron scattering spectrum of TGG at $T=5$~K measured by means of (a) IRIS, (b) and (c) MARI spectrometers with (a) PG (002) analyser, (b) $E_i=12$~meV and (c) $E_i=50$~meV. Intensities of inelastic features in all panels were normalised to 1. Polygons drawn with dashed line mark the integration ranges of data presented in Fig.~\ref{cef}.}
\label{cef_sqw}
\end{figure}

	The single ion ground-state of TGG is suggested to be a quasidoublet~\cite{leask94,nandi08}. The low energy part of the spectrum was investigated using IRIS, and at $T=5$~K, a single excitation at $E = 0.22$~meV can be seen (Fig.~\ref{cef_sqw} (a)). However, it is much broader than the instrumental resolution and has a strongly temperature dependent line shape that we will discuss further below. The first group of crystal field levels above the quasidoublet appears at $E\approx5$~meV, but individual levels cannot be resolved using either MARI with the best available resolution, or IRIS. The next features in the neutron spectrum are much weaker: at $E\approx28$~meV there are two distinct peaks; a few~meV above is another group at $E\approx35$~meV. An overview of all these features is shown in Fig.~\ref{cef_sqw} (b) and (c).

\begin{figure}
\centering
\includegraphics[width=\linewidth]{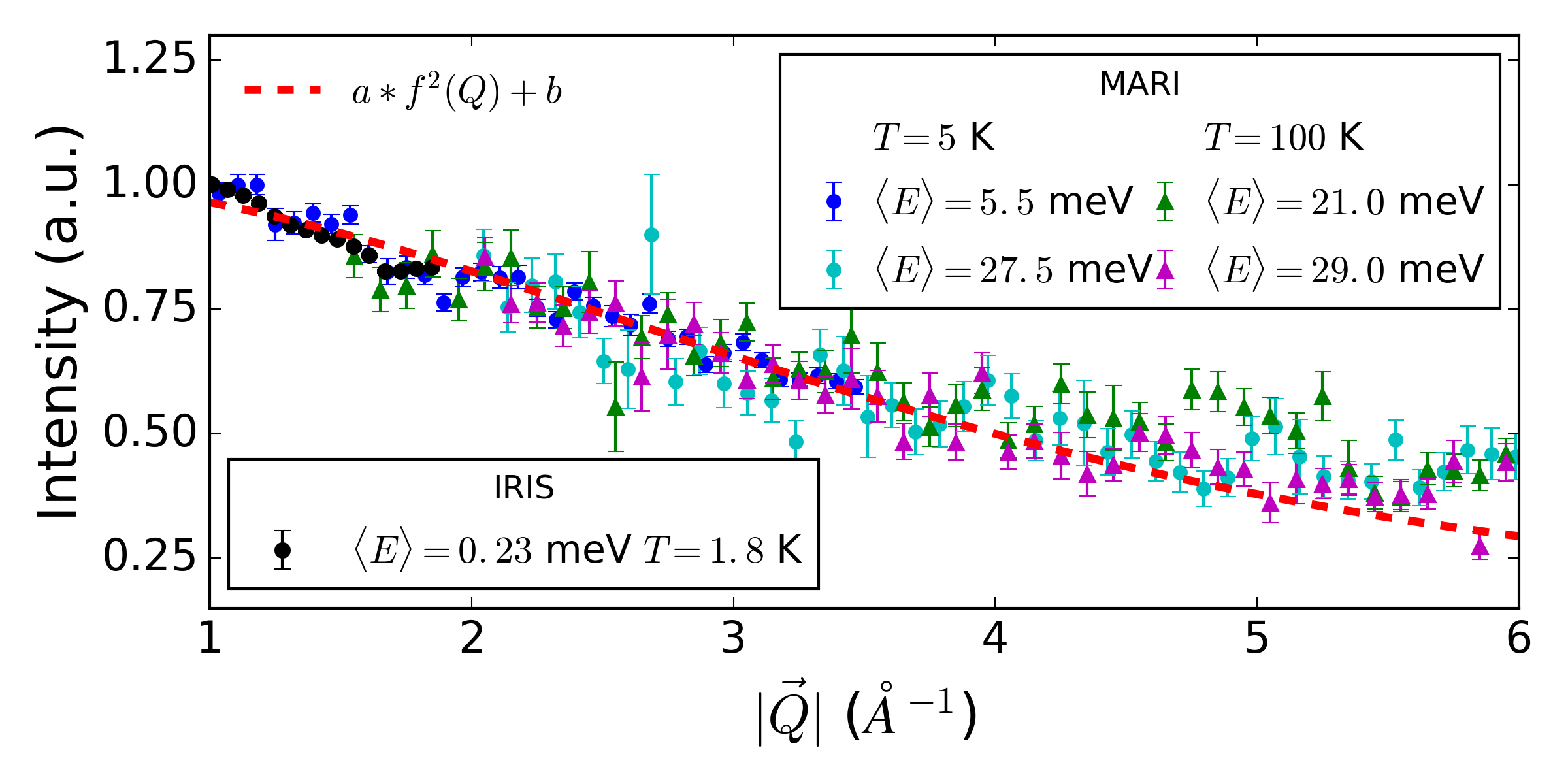}
\caption{$|\vec{Q}|$-dependence of the energy-integrated intensity of the observed features normalised to 1 at $|\vec{Q}|=1$. Blue and cyan markers show the transitions from ground-state measured at low temperature. Green markers show excited state transitions from levels at $E\sim{}5$~meV to the ones at $E\sim{}35$~meV that appear at $100$~K (Fig.~\ref{CEF_T_dep}). Black dots mark the quasidoublet transition intensity (Fig.~\ref{low_CEF_IRIS}). The red dashed line shows the magnetic form factor of Tb$^{3+}$ ion. The form factor is scaled and a small flat background is added to fit the low-$|\vec{Q}|$ data of low energy excitations (blue markers).}
\label{form_fac}
\end{figure}

	To ensure that all of the excitations we consider in our fitting are indeed crystal field excitations, their $|\vec{Q}|$-dependence was analysed. The energy-integrated intensity of the quasidoublet and the first two groups above it (i.e. $E\approx5,28$~meV) closely follow the dipole magnetic form factor of Tb$^{3+}$ ion~\cite{brown06}, as shown in Fig.~\ref{form_fac}. Phonon contributions make the resemblance to the calculated form factor less good in the high-$|\vec{Q}|$ region of the data. Retrieving a similar dependence from the $E\approx35$~meV feature shown in Fig.~\ref{cef_sqw}(c) was not possible due to the presence of an optical phonon branch at $E\approx40$~meV, but excited state transitions to these levels that follow the magnetic form factor were observed in the high temperature spectra, confirming their CEF nature and energies. The excitation that can be seen at $E\approx18$~meV in Fig.~\ref{cef_sqw}(c) was found to follow the $|\vec{Q}|^2$-dependence expected of a phonon branch. 

	To refine the parameters of the crystal field Hamiltonian, we require the energies of the crystal field excitations. Extracting these quantities is complicated when the separation of the excitations is close to the instrumental resolution, especially when the excitations are broadened beyond the resolution limit, and because of the close spacing of the two members of the quasidoublet. The energy difference of $\Delta=0.22$~meV between the two states of the quasidoublet means both levels are similarly populated at $T=5$~K ($F_1/F_0=e^{\frac{\Delta_{0\rightarrow1}}{kT}}=0.57$), and contributions to the spectra by excitations from both levels could be comparable.

\begin{figure}
\centering
\includegraphics[width=\linewidth]{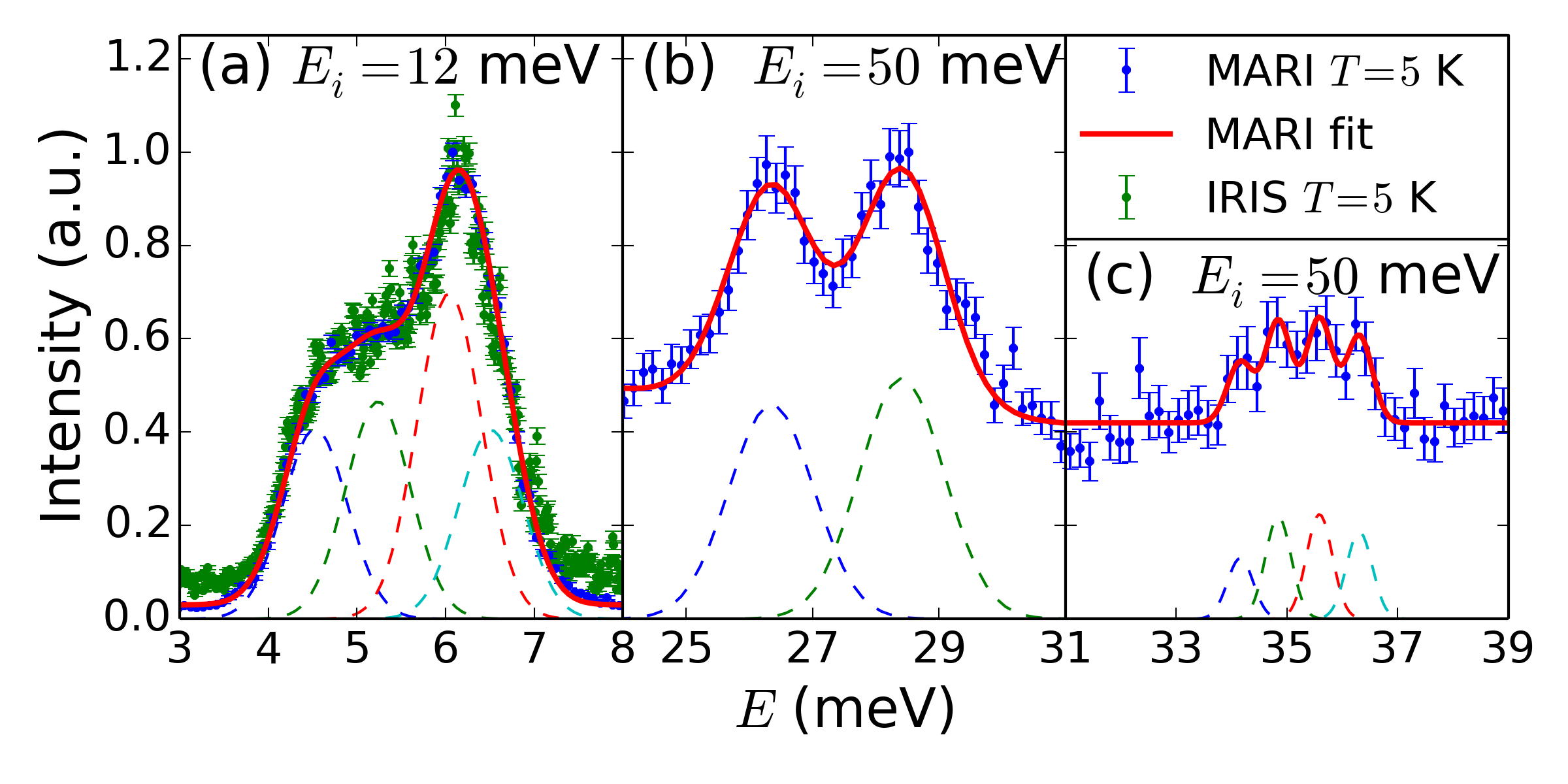}
\caption{Crystal-field transitions observed by inelastic neutron scattering with excitations fitted with Gaussian peaks. Data presented in figures is integrated around following average $|\vec{Q}|$-values: (a) $\langle|\vec{Q}|\rangle= 2.25$~\AA$^{-1}$, (b) $\langle|\vec{Q}|\rangle =4$~\AA$^{-1}$ and (c) $\langle|\vec{Q}|\rangle =3.5$~\AA$^{-1}$.}
\label{cef}
\end{figure}

	We integrated the data in ranges indicated in Fig.~\ref{cef_sqw}, using simple boxes for the levels at $E\approx5,28$~meV, and a parallelogram for the levels at $E\approx35$~meV. The resulting spectra are shown in Fig.~\ref{cef}. For the feature at $E\approx5$~meV (Fig.~\ref{cef}(a)), positions of single transition lines cannot be resolved using either spectrometer. Previous investigations of the TGG crystal field spectrum using Raman scattering~\cite{koningstein74} identified this feature as containing 4 excitations from the ground-state. We therefore fitted 4 Gaussian peaks constrained to have a common width ($FWHM=0.85(3)$~meV). A comparison between the reported transition energies~\cite{koningstein74} and the values deduced from the fits is made in Table \ref{cef_levels}, and shows good agreement between both datasets. At $E\approx28$~meV (Fig.~\ref{cef}(b)) two distinct peaks are observed, with a separation well exceeding the splitting of the ground-state singlets, suggesting they are separate levels. Both peaks were fitted with Gaussian peaks ($FWHM=1.93(7)$~meV). The group at $E\approx35$~meV (Fig.~\ref{cef}(c)) was fitted with four more Gaussians with $FWHM=0.55(5)$~meV, the minimum number of peaks required to obtain a good fit to this feature. We note that the widths of the measured features are considerably broader than the instrumental resolution, with the exception of the members of the group lying around $35$~meV.

\begin{figure*}
\centering
\includegraphics[width=\linewidth]{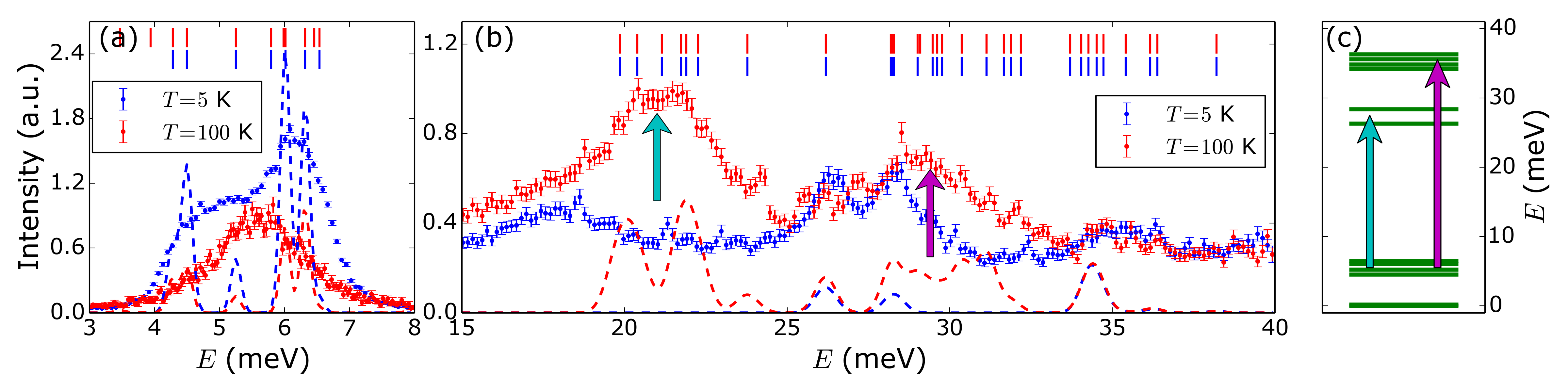}
\caption{(a) $|\vec{Q}|$-integrated ($\langle|\vec{Q}|\rangle =2.25\pm{}1.25$~\AA$^{-1}$) excitation spectra measured on MARI with $E_i=12$~meV. (b) $|\vec{Q}|$-integrated ($\langle|\vec{Q}|\rangle =4\pm{}2$~\AA$^{-1}$) excitation spectra measured at MARI with $E_i=50$~meV. At $T=100$~K two features appear with $|\vec{Q}|$-dependence following the magnetic form factor (Fig.~\ref{form_fac}). Their energy correspond to the transitions, represented by cyan and magenta arrows, in the schematic CEF excitation spectrum shown in (c). Dashed lines present the neutron scattering intensities calculated within dipole approximation convoluted with Gaussian peaks. For illustrative purposes we use the FWHM retrieved from resolution calculations for MARI with proper chopper setup. Vertical lines mark the energies of expected transitions from thermally populated states having non-zero intensities.}
\label{CEF_T_dep}
\end{figure*}

	Changes in the measured spectra between $T=5$~K and $T=100$~K are shown in Fig.~\ref{CEF_T_dep}(a) and (b). The most striking difference is the appearance of strong intensities at $E\approx21.29$~meV, which follow the magnetic form factor of Tb$^{3+}$ (Fig.~\ref{form_fac}), and can therefore be identified as transitions from thermally populated levels at $E\approx5$~meV (\ref{CEF_T_dep}(c)), consistent with the assignment of excitations at $E\approx35$~meV, whose form factor could not be verified, as a crystal field excitation. 
	
\begin{table}
\caption{CEF transition energies observed by Raman spectroscopy~\cite{koningstein74}, calculated with parameters given in Ref.~[\onlinecite{guillot85}], determined by fits to INS data (Fig.~\ref{cef}(a), (b) and (c)) and calculated with the refined parameters (Tab. \ref{cef_params}). All values are given in~meV. \label{cef_levels}}
\begin{ruledtabular}
\begin{tabular}[c]{ccccccc}
$E_{obs Raman}$\cite{koningstein74}	& $E_{calc}$\cite{guillot85}	& $E_{obs INS}$	& $E_{calc}$		\\	\hline
-								& 0.4						& 0.219(1)			& 0.22			\\				
4.2								& 4.8						& 4.52(2)			& 4.50			\\
5.3								& 5.0						& 5.23(3)			& 5.25			\\
6.2								& 5.3						& 6.04(4)			& 6.01			\\
6.6						 		& 6.1						& 6.51(3)			& 6.54			\\
-								& 36.6					& 26.30(4)			& 26.40			\\
-								& 38.6					& 28.38(5)			& 28.27			\\
-								& 44.6					& 34.16(8)			& 34.26			\\
-								& 46.7					& 34.85(5)			& 34.73			\\
-								& 49.9					& 35.59(5)			& 35.63			\\
-								& 50.7					& 36.31(5)			& 36.38			\\
-								& 55.4					& -				& 38.20			\\
\end{tabular}
\end{ruledtabular}
\end{table}
		
\subsubsection{Crystal Electric Field Hamiltonian}
	The crystal-field Hamiltonian for $D_{2}$ point-group symmetry is given by
\begin{equation}
\label{eq:cef_H_Stevens}
\begin{split}
\hat{\mathcal{H}}_{\rm CEF} 	
& = \widetilde{B}_0^2\hat{O}_0^2+\widetilde{B}_2^2\hat{O}_2^2					\\
& + \widetilde{B}_0^4\hat{O}_0^4+\widetilde{B}_2^4\hat{O}_2^4+\widetilde{B}_4^4\hat{O}_4^4	\\
& + \widetilde{B}_0^6\hat{O}_0^6+\widetilde{B}_2^6\hat{O}_2^6+\widetilde{B}_4^6\hat{O}_4^6 + \widetilde{B}_6^6\hat{O}_6^6,
\end{split}
\end{equation}
where $\hat{O}_{q}^{k}$ are the (cosine) Stevens operators $\hat{O}_{q}^{k}(c) \equiv \hat{O}_{q}^{k}(\hat{J}_{z},\hat{J}_{\pm})$. The crystal-field coefficients $\widetilde{B}_q^k= \langle r^k \rangle \theta_k A_q^k(c) $ can be found in standard literature about the Stevens' equivalent method~\cite{hutchings64,abragambleaney70,jensen91}. 

	We can summarise the general {expressions} of the CEF eigenstates as\begin{equation}
\begin{split}
\ket{\Gamma_1} =& \sum_{M_J=2,4,6} b_{M_J} \left( \ket{M_J} + \ket{-M_J} \right)	+	b_{0}	\ket{0},	\\
\ket{\Gamma_2} =& \sum_{M_J=1,3,5} a'_{M_J} \left( \ket{M_J} - \ket{-M_J} \right)	,						\\
\ket{\Gamma_3} =& \sum_{M_J=1,3,5} a_{M_J} \left( \ket{M_J} + \ket{-M_J} \right),							\\
\ket{\Gamma_4} =& \sum_{M_J=2,4,6} b'_{M_J} \left( \ket{M_J} - \ket{-M_J} \right),
\end{split}
\label{eq:J6statesMJ}
\end{equation}
where all coefficients $a_{M_J}, b_{M_J}, a'_{M_J}, b'_{M_J}$ are real, and the $\ket{M_J} \equiv \ket{J=6, M_{J}}$ are eigenstates of the $\hat{J}_{z}$ operator.
The generic states $\ket{\Gamma_{\alpha}}$ are labelled according to their irreducible representation $\Gamma_{\alpha}$, which we deduce by comparison with character tables.

	Previous analyses of isostructural Tb$_{3}$Al$_{5}$O$_{12}$ (TAG) have attempted to simplify the CEF scheme by retaining cubic or tetragonal symmetries (reducing the number of operators to 4 or 5 respectively), with the orthorhombic terms expected to produce only a small modification of the parameters. We find that such approaches cannot describe our experimental data, and so we use the orthorhombic Hamiltonian throughout the following. This is in contrast to other RE garnets (see for example Refs.~[\onlinecite{gavignet-tillard73,reid00}]) where the use of higher symmetries has been successful.

	Refinement of the 9 crystal-field parameters $\widetilde{B}^k_q$ of Eq.~\ref{eq:cef_H_Stevens} was attempted with the use of the \texttt{SPECTRE} software~\cite{spectre}. The refined quantity was the transition energies (and not the intensities), retrieved from fitting to the INS spectra. Refinement of values previously published for TGG~\cite{guillot85}, from the results of Raman spectroscopy measurements of a small number of levels (see Table \ref{cef_levels}) and single crystal magnetisation, did not lead to satisfactory results. Several other sets of initial parameters were tested (Dy~\cite{crosswhite67}, Er~\cite{orlich69}, Eu~\cite{boal73} and Nd~\cite{antic-fidancev86} gallium garnets scaled accordingly to the differences between radial momenta of RE$^{3+}$ and Tb$^{3+}$ ions, and parameters for Tb ions in YAG~\cite{morrison76}). We have made attempts of refinement with sets of basis functions limited to the $^7F_6$ ground-state multiplet ($LS$-coupling), $^7F_N$ ($N\in{}\langle0,6\rangle$) states lying between $253$ and $698$~meV~\cite{bayerer86} (intermediate coupling), and including all higher $^{2S+1}L_J$ terms. Although good agreement between calculated and observed energies was found in the two latter cases (twofold reduction of goodness of fit parameter when going beyond $LS$ coupling and without any further improvement when expanding the basis more), a proper convergence of this multi-parameter refinement was not achieved. The previously listed initial sets of parameters together with some of the best final sets refined with \texttt{SPECTRE} were introduced into a custom least-squares fitting routine refining the eigenvalues of the Stevens Hamiltonian (Eq. \ref{eq:cef_H_Stevens}) (i.e. restricted to $LS$-coupling). The algorithm reached convergence and the results ranged from good to excellent agreement with the measured values. The optimal solution was chosen by investigating the neutron spectra calculated using the refined parameters and comparing them with INS data (Fig.~\ref{CEF_T_dep}) and the set giving the closet distribution of calculated intensities with respect to the observed one was chosen (based on the result of the initial EuGG refinement with \texttt{SPECTRE}). This result was also characterised by the best value of the goodness of fit parameter $\chi^2=2.1\times{}10^{-3}$. The contributions to the wavefunctions in the $|M_J \rangle$ basis in Table \ref{eigenvectors} is given in Appendix \ref{eigenvectors_sec}.

\begin{table}
\caption{Set of refined crystal-field parameters giving the best value of standard goodness of fit parameter, $\chi^2=2.1\times{}10^{-3}$. Parameters are given in Stevens notation. \label{cef_params}}
\begin{ruledtabular}
\begin{tabular}[c]{ccccccc}
Parameter	& Ref.~[\onlinecite{guillot85}] (meV)	& This work (meV)					\\	\hline
$\widetilde{B}^0_2$		& $5.05\times10^{-2}$			& $2.66\times10^{-2}$			\\
$\widetilde{B}^2_2$		& $-2.60\times10^{-1}$			& $-3.24\times10^{-1}$			\\
$\widetilde{B}^0_4$		& $-4.10\times10^{-3}$			& $-2.47\times10^{-3}$			\\
$\widetilde{B}^2_4$		& $3.00\times10^{-3}$			& $3.70\times10^{-3}$			\\
$\widetilde{B}^4_4$		& $1.50\times10^{-2}$			& $8.51\times10^{-3}$			\\
$\widetilde{B}^0_6$		& $-5.89\times10^{-6}$			& $-1.12\times10^{-5}$			\\
$\widetilde{B}^2_6$		& $1.36\times10^{-5}$			& $-1.21\times10^{-5}$			\\
$\widetilde{B}^4_6$		& $-1.02\times10^{-4}$			& $-4.73\times10^{-5}$			\\
$\widetilde{B}^6_6$		& $5.32\times10^{-7}$			& $8.96\times10^{-5}$			\\
\end{tabular}
\end{ruledtabular}
\end{table}
	
	The neutron scattering intensities calculated within the dipole approximation using refined CEF parameters for $T=5$~K and $T=100$~K are plotted along with experimental results in Fig.~\ref{CEF_T_dep}(a) and (b). The calculated positions of the excitations are in good agreement with measured intensities. The shapes of the features at higher energies are reasonably reproduced by the calculations. In addition, the almost non-existent scattering intensity calculated for the transition between the ground-state and the highest state of the lowest multiplet is consistent with it not being observed in the experimental data.
	
	When compared with previous determinations of the crystal field parameters, i.e. Ref.~[\onlinecite{guillot85}] (see Table~\ref{cef_params}), we find reasonable agreement between most parts of the Stevens parameters, but there are differences in the size and sign, particularly for the highest order parameters. As can be seen in Table~\ref{cef_levels}, rather few of the levels were previously determined experimentally, with the results that the parameters are not very accurate. Indeed, when the full level scheme is calculated using the parameters of Ref.~[\onlinecite{guillot85}], we can see that the higher levels are predicted at much higher energies than we observe them (see Table~\ref{cef_levels} for comparison). 

	Nonetheless, these previously determined parameters and eigenfunctions could be used successfully to reproduce a step-like magnetisation curve measured on a single crystal sample at $4.2$~K~\cite{guillot85}. However, all of the parameter sets obtained in the course of our refinements give a similar structure of the wavefunctions. They also describe the magnetisation step, which is a rather generic feature present for similar compositions of the quasidoublet eigenstates, and one of the levels at $E_n\sim5$ meV. The significant contribution from high-$M_J$ states to this higher level causes, due to the Zeeman term, its rapid descent in field towards the energies of the quasidoublet states leading to increase in its thermal population and contribution to the overall magnetisation. By comparison of the quasidoublet wavefunctions obtained from our parameter set to the one obtained from the parameter set in Ref.~[\onlinecite{guillot85}], we find that the symmetry character $|\Gamma_3\rangle$ and $|\Gamma_1\rangle$, of the ground singlet, $|p \rangle \equiv |\psi_{0}\rangle$, and the first-excited one, $|q \rangle \equiv |\psi_{1}\rangle$, are swapped. More explicitly, $|p \rangle \equiv |\psi_{0}\rangle \approx­0.09|\pm5\rangle­0.03|\pm3\rangle+0.7|\pm1\rangle$ and $|q \rangle \equiv |\psi_{1}\rangle\approx0.18|\pm6\rangle+0.12|\pm4\rangle­0.33|\pm2\rangle­0.83|0\rangle$ in this work [Tab.~\ref{eigenvectors} in Appendix \ref{eigenvectors_sec}], whereas in Ref.~[\onlinecite{guillot85}] we have $|p \rangle \approx­0.05|\pm6\rangle+0.17|\pm4\rangle­0.16|\pm2\rangle­ 0.94|0\rangle$ and $|q \rangle \approx­0.06|\pm5\rangle­0.15|\pm3\rangle+0.69|\pm1\rangle$. The order of the $|\Gamma_3\rangle/|\Gamma_1\rangle$ states was found to play an important role in obtaining a good reproduction of the transition intensities in INS.
	
\begin{figure}
\centering
\includegraphics[width=\linewidth]{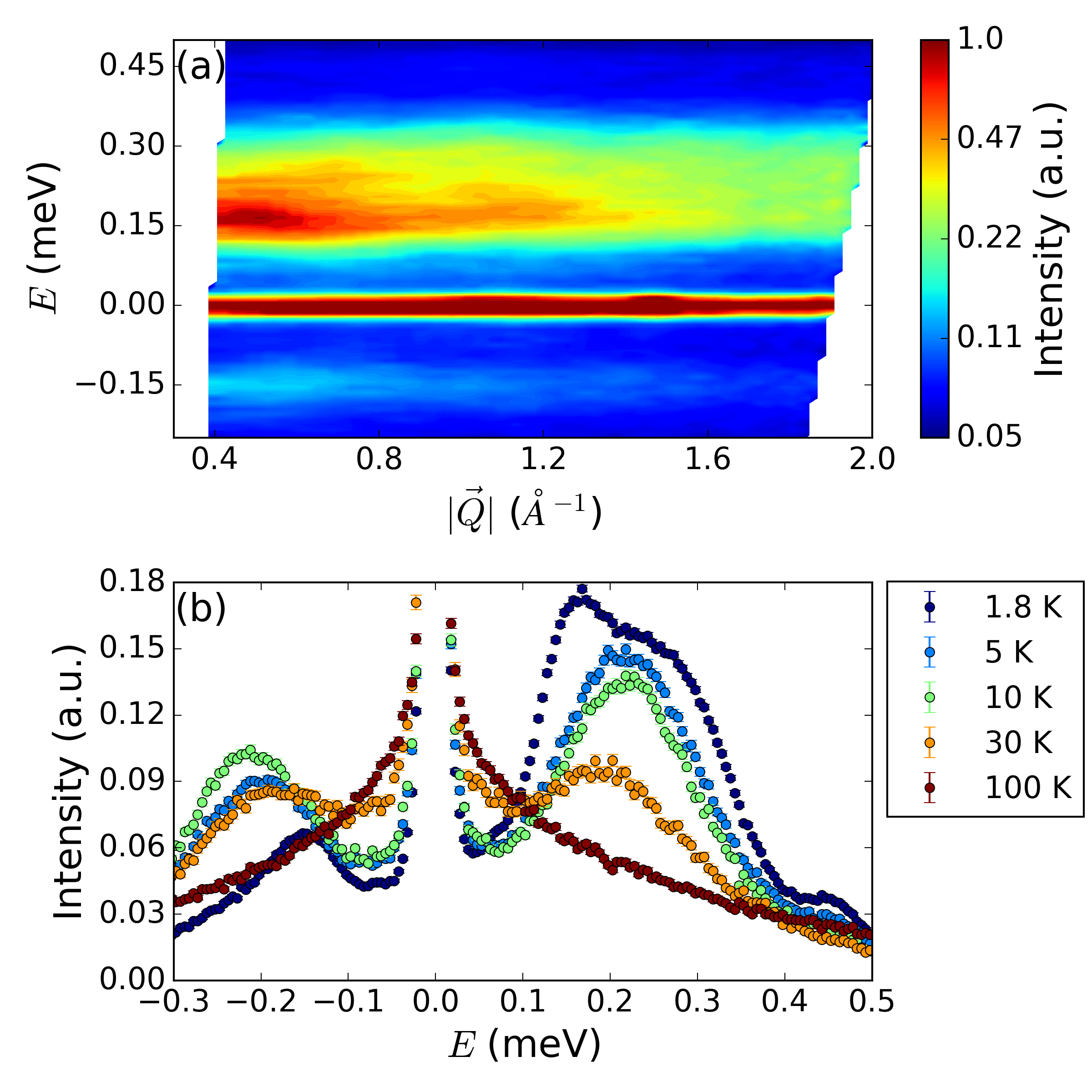}
\caption{(a) $S(|\vec{Q}|,\omega)$ of the low-lying CEF singlet measured on IRIS at $T=1.8$~K. Hints of dispersive character of the measured excitation can be observed. Intensity of inelastic feature normalized to 1 on logarithmic scale. (b) Temperature dependence of $|\vec{Q}|$-integrated ($|\vec{Q}|$=0.4-1.8~\AA$^{-1}$) cuts through the ground-state quasidoublet.}
\label{low_CEF_IRIS}
\end{figure}

	On the other hand, at $E\approx5$~meV, there is a strong discrepancy between the calculated and observed line widths, which are much larger than the instrumental resolution. Furthermore, although a single transition between the members of the quasidoublet can be seen at $E=0.22$~meV at $T=5$~K (though the peak is again much broader than the instrumental resolution), the lowest energy part of the spectrum displays a very strong temperature dependence. At the lowest measured temperature $T=1.8$~K the $S(|\vec{Q}|,\omega{})$ map of this part of the spectrum [Fig.~\ref{low_CEF_IRIS}(a)] suggests that the single excitation is replaced by a band of dispersive excitons. A clearly visible contribution of two or more peaks to the first transition has developed, as shown in Fig.~\ref{low_CEF_IRIS}. On the high energy tail of the main peak is another weak peak at $E\approx0.45$~meV that develops with the same temperature dependence as the multi­peak structure of the main excitation. A shoulder can be distinguished in the quasidoublet gap at $T<3.5$~K, which suggests another peak. This observation, along with the large widths of other excitations (i.e. $|\psi_n\rangle$ with $E_n\sim{}5$~meV) , suggests interactions between Tb$^{3+}$ ions, or coupling with lattice excitations modifies the excitation spectrum beyond a single-ion picture.
	
\subsection{Magnetisation}\label{sec:mag}
\begin{figure}
\centering
\includegraphics[width=\linewidth]{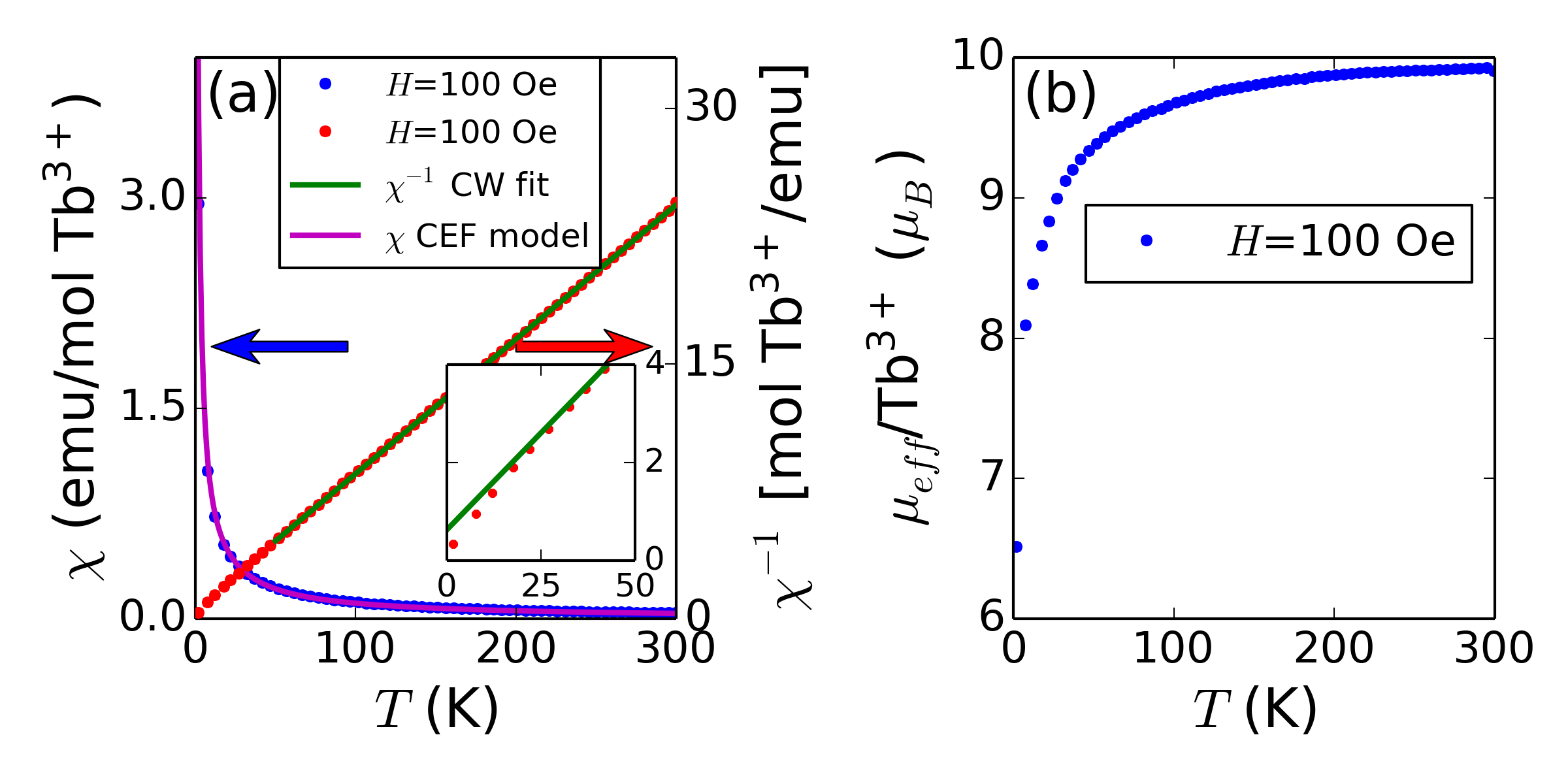}
\caption{(a) Powder magnetisation $M$ as a function of temperature $T$ plotted in form of $\chi(T)\sim{}M(T)/H$ and $\chi^{-1}(T)$. Solid magenta line shows susceptibility calculated with use of refined CEF parameters and solid green line shows fit to the Curie-Weiss law with $\Theta_{CW}=-7.98$~K. (b) Effective paramagnetic moment per Tb$^{3+}$ ion $\mu_{\rm eff}=\sqrt{\frac{3 k_B}{N_A\mu_{\mathrm{B}}^2}\chi T}$.}
\label{chi}
\end{figure}

	The magnetic susceptibility is presented in Fig.~\ref{chi}(a). Above $T=50$~K it can be fitted by a Curie-Weiss law with $\theta_{CW}=-7.98$~K (hence the frustration parameter $f\equiv{}|\theta_{CW}|/T_{\rm N}\sim{}32$). 
By means of $\mu_{\rm eff}=\sqrt{\frac{3k_B}{N_A\mu_{\mathrm{B}}^2}\chi T}$, we extracted $\mu_{\rm eff}=9.913(3)$ $\mu_{\mathrm{B}}$ as paramagnetic moment per Tb$^{3+}$ ion over the range $260<T<300$~K. Such value differs by only 2\% from $\mu_{\rm Tb^{3+}}=9.72$ $\mu_{\mathrm{B}}$, the theoretical paramagnetic moment $\mu=|\mathbf{m}|= g_{J} \sqrt{J(J+1)}$ $\mu_{\mathrm{B}}$ per Tb$^{3+}$ ion in the ${^7F_6}$ ground multiplet~\cite{jensen91}. 
Calculation of the susceptibility using the refined crystal-field parameters (Tab. \ref{cef_params}) gives excellent agreement with the experimental data.
	
\begin{figure*}
\centering
\includegraphics[width=\linewidth]{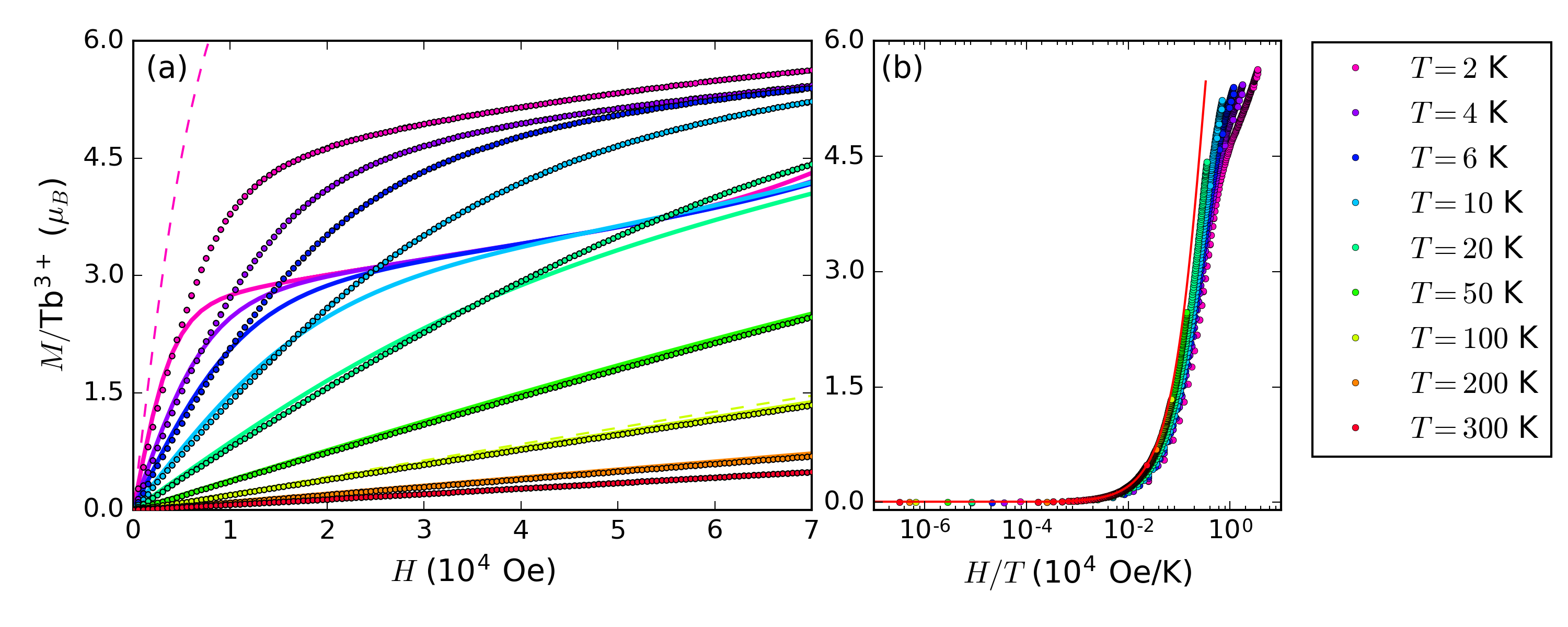}
\caption{Isothermal powder magnetisation curves as a function of (a) field $H$ (b) and $H/T$ ratio. Low-$T$ curves in (b) show the departure from scaling expected for the system of uncorrelated spins. Solid and dashed lines in (a) are: results of calculations performed with refined CEF parameters and Brillouin functions calculated for the isotropic case of free Tb$^{3+}$ ion, respectively. The solid red line in (b) marks the expected master curve calculated from CEF scheme for $T=300$~K.}
\label{M_H}
\end{figure*}
	
	The magnetisation per Tb$^{3+}$ ion as a function of applied field, $H$, for different temperatures, and in terms of the ratio $H/T$ is plotted in Fig.~\ref{M_H}(a) and \ref{M_H}(b), respectively. Solid lines in panel (a), show the result of calculations using the refined CEF parameters (Tab. \ref{cef_params}), which are in good agreement with the data down to 20~K. Below this temperature, the calculated results deviate strongly from the experimental data. Similarly, the collapse of the $M(H/T)$ curves that is expected for paramagnetic spins, can be observed down to 50~K. For $T<50$~K deviation from this scaling, coinciding with the departure of the susceptibility from the Curie-Weiss law, becomes increasingly pronounced (inset of Fig.~\ref{chi}(a)). These behaviours are consistent with the onset of a correlated cooperative paramagnetic phase. Our measured susceptibility and magnetisation agree well with another recent report~\cite{Mukherje2017}.

\section{Discussion}

	We have investigated the crystal and magnetic structures of TGG, together with the crystal field Hamiltonian for its single-ion physics. These are all relevant for its low temperature physics and magnetothermal effects. 

	Firstly, the crystal structure of our samples is close to the ideal garnet structure, but there are superstoichiometric terbium ions (Tb2). These Tb$^{3+}$ ions in excess act as defects which enhance the scattering of phonons and cause the lowering of the thermal conductivity~\cite{inyushkin07}. For this reason, in Czochralski-grown crystals, where their abundance goes up to $\approx{}1$ \%, they are considered responsible for the manifestation of the thermal Hall effect~\cite{inyushkin10}. Such concurrence -- Czochralski-grown crystals have the highest percentage of Tb2 defects and lowest diagonal thermal conductivity -- motivated the theoretical work of Ref.~[\onlinecite{mori14}], in which Mori \emph{et~al.} posit that the resonant skew-scattering of phonons from the CEF states of superstoichiometric Tb$^{3+}$ ions is the \emph{origin} of the thermal Hall effect. In this theory two assumptions are crucial: (i) the Tb2 ions are somewhat correlated in space, (ii) the CEF splitting of their lowest levels (labelled as $|a\rangle, |b\rangle$ in Ref.~[\onlinecite{mori14}]) is comparable with $\Delta$, the CEF gap for the quasidoublet states, $|p\rangle, |q\rangle$, of the regular Tb sites. 

	Our study provides insights to this theory, from both structural and single-ion perspectives. We detect superstoichiometric terbium ions only on the 16a (Ga1) Wyckoff position, where their concentration is 1.6~\%. The importance of this observation is that the $16a$ position has a $C_{3i}$ symmetry, which is higher than the $D_2$ and $S_4$ symmetries, respectively for the $24c$ (Tb) and $24d$ (Ga2) sites (see Tab.~\ref{cryst_param}). For $J=6$, 5 singlets and 4 doublets are expected in $C_{3i}$ symmetry, and the assumption that the crystal field scheme at this site will be generally similar to that at the main terbium site is to be examined carefully. Results from point charge calculations (which include the 6 nearest oxygen ions) suggest a similar structure of the quasidoublet ground state with 4 symmetry-allowed doublets lying at energies higher than $90$~meV. However, one has to keep in mind the limited reliability of this method. Regarding the spatial correlations of Tb2 ions and their crystal field scheme, direct experimental probing is hindered by the low percentage of superstoichiometric abundance in TGG. 

	The low-temperature magnetic structure of TGG that we have observed is a multiaxial antiferromagnet composed of magnetic moments that have non-collinear local anisotropy axes. Any singlet $\ket{\psi_{n}}$, from the CEF spectrum of the Tb$^{3+}$ ion in TGG, is characterised by $\braket{ \psi_{n} | \hat{\mathbf{J}} | \psi_{n} }=0$ ($\hat{\mathbf{J}} \equiv (\hat{J}_{x} , \hat{J}_{y}, \hat{J}_{z})$ are the angular momentum operators), a consequence of the complete removal of the degeneracy of the free ion states by the crystal field~\cite{griffith63,abragambleaney70}. The magnetic moments found in the ordered phase of TGG must therefore be induced by perturbations which couple or mix the CEF ground singlet with higher CEF eigenstates, i.e. an induced moment system. Usually this mixing is achieved by the exchange interactions, but a curious feature of the magnetic structure is that it nullifies the nearest neighbor Heisenberg terms $ \textbf{S}_i \cdot \textbf{S}_j$, since the spin expectation values of induced moments $ \textbf{S}_i$ are expected to be orthogonal (see below Eq.~\eqref{eq:MagneticMomentDirections}). Our measurements of the magnetic susceptibility and magnetization suggest nonetheless that antiferromagnetic correlations build up below $T\sim{}20$~K, far above the magnetic ordering temperature. One may well ask how this combination of anisotropy, frustration, and magnetic order arises. 

	An insightful general argument {about anisotropies in non-Kramers systems} was given by Griffith in Refs.~[\onlinecite{griffith63}], and it concerns instances where an even number of low-lying singlets is well isolated from higher excited states. In the simplest case of two low-lying singlets it was shown that the axial anisotropy of the magnetic moment is determined by their time-reversal (TR) behaviour, i.e. the local Ising axis of a non-Kramers ion is independent of the direction of the applied field (provided this is only of perturbative order~\cite{Note1}). 

	As already pointed out in Ref.~[\onlinecite{bidaux73}], this is of interest for the quasidoublet states ($\ket{p}, \ket{q}$) of the CEF in similar garnets and, in order to understand and possibly predict which of the three two-fold axes of the $D_{2}$ CEF-symmetry is the anisotropy axis for the magnetic moment, we follow Ref.~[\onlinecite{griffith63}] and investigate the matrix elements $\braket{q| \hat{\mathbf{J}} |p}$ and their behaviour under TR transformation.

	The action of the time-reversal operator $\Theta$ on angular momentum states $\ket{\psi} = \sum_{J,M_J}	C_{J,M_J} \ket{J,M_J}$ is given by $\ket{\psi}^{K}=\Theta \ket{\psi}= \sum_{J,M_J} C^{*}_{J,M_J}(-1)^{J-M_J} \ket{J,-M_J}$~\cite{abragambleaney70}. Then the time-reversal of the CEF states in Eq.~\eqref{eq:J6statesMJ} reads	\begin{equation}
\begin{split}
\ket{\Gamma_1}^{K} =& \ket{\Gamma_1},	\\
\ket{\Gamma_2}^{K} =& \ket{\Gamma_2},	\\
\ket{\Gamma_3}^{K} =& -	\ket{\Gamma_3},	\\
\ket{\Gamma_4}^{K} =& -	\ket{\Gamma_4}.
\end{split}
\label{eq:TimeReversalJ6statesMJ}
\end{equation}
Calculating the matrix elements $\braket{ \Gamma_{\alpha} | \hat{\mathbf{J}} | \Gamma_{\beta} }$, by means of the general expansions in Eq.~\eqref{eq:J6statesMJ}, reveals the correspondence \begin{equation}
\begin{split}
\mathbf{m}({ \Gamma_3 , \Gamma_1 }	)	\parallel	\mathbf{m}({ \Gamma_2 , \Gamma_4 })	&	\parallel	\mathbf{x}_{i},	\\
\mathbf{m}({ \Gamma_3 , \Gamma_4 }	)\parallel	\mathbf{m}({ \Gamma_2 , \Gamma_1 })	&	\parallel	\mathbf{y}_{i},		\\
\mathbf{m}({ \Gamma_3 , \Gamma_2 })	\parallel	\mathbf{m}({ \Gamma_1 , \Gamma_4 })	&	\parallel	\mathbf{z}_{i},
\end{split}
\label{eq:MagneticMomentDirections}
\end{equation}
between the symmetry of the states and the direction of $\mathbf{m}(\Gamma_{\alpha}, \Gamma_{\beta})= ­- g_{J} \mu_{\mathrm{B}} \braket{ \Gamma_{\alpha} | \hat{\mathbf{J}} | \Gamma_{\beta}}$, being the (``off-diagonal'', $\alpha \neq \beta$) components of the (induced) magnetic moment in TGG. From Eq.(\ref{eq:J6statesMJ}-\ref{eq:MagneticMomentDirections}), we deduce that off-diagonal matrix elements predict a magnetic moment pointing along the local $\mathbf{x}_{i}, \mathbf{y}_{i}, \mathbf{z}_{i}$ axes depending on the $M_J$-decomposition and time-reversal (TR) properties of the two states of the quasidoublet. More specifically we find the induced moment $\mathbf{m}(\Gamma_{\alpha}, \Gamma_{\beta})$ along $\mathbf{x}_{i}$ if $ \ket{\Gamma_{\alpha}},\ket{\Gamma_{\alpha \neq \beta}}$ have different basis decomposition (one with $M_J$-even and the other with $M_J$-odd) and different TR behaviour (one even the other odd), along $\mathbf{y}_{i}$ if they have different basis decomposition but same TR transformation, and along $\mathbf{z}_{i} $ if they have same basis decomposition but opposite TR transformation. 

	Diagonalisation of the crystal-field Hamiltonian in Eq.~\eqref{eq:cef_H_Stevens} with the parameters listed in Tab.~\ref{cef_params} leads to a spectrum where the lowest singlet $|p \rangle$ and the first excited one $|q\rangle$ are, respectively, of the $\ket{\Gamma_3}$ and $\ket{\Gamma_1}$ type in Eq.~\eqref{eq:J6statesMJ} {(see also Tab.~\ref{eigenvectors})}, $\mathbf{m}_{i}(p, q) \parallel \mathbf{x}_{i}$, according to Eq.~\eqref{eq:MagneticMomentDirections}, and as observed in the magnetic structure. We find indeed that $\braket{p|\hat{J}_{y}|q}=\braket{p|\hat{J}_{z}|q}=0$, identically, and so the low-temperature induced magnetic moment will be proportional to $\mathbf{m}_{i}(p, q)= - g_{J} \mu_{\mathrm{B}} \braket{p|\hat{J}_{x}|q} \mathbf{x}_{i}$ (see Appendix~\ref{sec:coordinates} for the different $\mathbf{x}_{i}$ in TGG).

	The magnetic structure of TGG that we have determined is identical to that found previously~\cite{hammann75}, and to that found in various other rare earth garnets, including Dy$_3$Al$_5$O$_{12}$ (DAG). In the case of DAG, Wolf and collaborators showed that this magnetic structure minimises the dipolar interactions for magnetic moments with $\langle 100\rangle$ Ising anisotropy and overall constraint of cubic symmetry\cite{ball63}, and this was subsequently confirmed by the application of the Luttinger-Tisza method~\cite{felsteiner81}. In analogy with these RE-garnets, it is typical also in TGG to ascribe the ordered structure predominantly to dipolar interactions.

In the theoretical description of the ordering transition in TGG of Ref.~[\onlinecite{hammann73}], Hammann and Manneville also concluded that the dipolar interaction is of primary importance, but found that the observed ordering temperature was too high to be accounted for only by induced moment ordering mediated by dipolar interactions~\cite{hammann73}. They therefore showed that a molecular field theory taking into account also the role of hyperfine coupling, together with the dipolar interactions, would provide a reasonable estimate of $T^{\rm dip}_{\rm N} \approx0.345$~K with respect to $T^{\rm exp}_{\rm N}\approx0.25$~K, and implying the possibility of cooperative electronic and nuclear spin order. This theory was constrained by various experimental parameters of the time, which can now be examined more accurately. 

	Our study confirms directly the value of the (ground state) quasidoublet gap, and it suggests a time-reversal analysis of the direction of the local easy axes (among the three equivalent ones permitted by the crystalline point symmetry). The afore-mentioned ``off-diagonal'' matrix element $\braket{p|\hat{J}_{x}|q} \approx5.3$, is a crucial quantity which relates to the coefficient $w$ in Ref.~[\onlinecite{hammann73}] \emph{via} $|w|= \left| \mathbf{m}(p, q)\right|$, and therefore controls the key parameters for the mean-field theory presented therein (see Tab.~\ref{tab:HammanQuantities}). It is worth emphasising that our estimation ($w \approx 7.9 \, \mu_{\mathrm{B}}$ ) is given by the wave functions, $\ket{p}, \ket{q}$, of the ground-state quasidoublet obtained \emph{via} exact diagonalisation of the Hamiltonian in Eq.~\eqref{eq:cef_H_Stevens}, i.e. crystal-field analysis of the neutron scattering data. In contrast, the estimation by Hammann and Manneville ($w \approx 6.7 \, \mu_{\mathrm{B}}$) relies on a self-consistent fit to magnetisation curves measured at different temperatures on a single crystal sample with magnetic field applied along the $\langle111\rangle$ direction~\cite{Note3}.

\begin{table}
\def\arraystretch{1.5}
\caption{Estimation of the key parameters for the mean-field theory in Ref.~[\onlinecite{hammann73}]. The quantities are given for Tb$^{3+}$ ions, where the Land\'e factor is $g_{\rm J}=3/2$ and the hyperfine coupling constant is $a_{\rm J} \approx2.2 \, 10^{-3}$~meV~\cite{hammann73} (see Table 5.5 in Abragam and Bleaney~\cite{abragambleaney70}).}
\label{tab:HammanQuantities}
\begin{ruledtabular}
\begin{tabular}[c]{ccccccc}
							& Ref.~[\onlinecite{hammann73}] 	& This work 	\\	\hline
$|w| / \mu_{\mathrm{B}}$		&			6.7					& 7.9			\\	
$|\alpha|$					&			0.08					& 0.11	
\end{tabular}
\end{ruledtabular}
\end{table}

	Our direct estimation of $w$ opens up a possible route to understand coherently the behaviour of the magnetic moment across different temperature regimes. In a forthcoming theoretical work we will study how the local states ($\ket{p}, \ket{q}$) of the Tb$^{3+}$ ions are the starting point for the definition of the new magnetic basis  ($\ket{p'}, \ket{q'}$) which carries information about the admixture induced by the presence of the (molecular) field on one hand and of the thermal population of states on the other. (A similar reasoning was considered in Refs.~[\onlinecite{bidaux73}] and [\onlinecite{hammann73}], but actual local symmetries of the RE$^{3+}$ ions were disregarded as an element of interest.) This route can resolve the puzzle reported in Ref.~[\onlinecite{hammann75}], where neutron powder diffraction showed that full saturation of the magnetic moment was still not achieved down to 0.2~$T/T_{\rm N}$ (at such temperatures $\left| \mathbf{m}\right|/w\approx0.6$ where $w$ is treated as the saturated magnetic moment per ion). In addition, this is expected to be the natural starting point to investigate which mechanism among others -- to mention a few, competing interactions, fragmentation, fluctuations (outside of the neutron spectroscopy window) -- act as leading term in the reduction of the magnetic moment.

	Following the perspective of Ref.~[\onlinecite{hammann73}], the key parameters in Tab.~\ref{tab:HammanQuantities} also comprehend implications about the role of hyperfine fields in the magnetic ordering of TGG. The coefficient $ |\alpha| = 2 |w| a_{\rm J}/ \Delta g_{\rm J} \mu_{\mathrm{B}}$, is the dimensionless strength of the hyperfine coupling in the pseudospin-$1/2$ effective Hamiltonian obtained by projecting on the quasidoublet subspace -- here $\Delta$ is the CEF ground state splitting (see Sec.~\ref{sec:local}) and $a_{\rm J}$ comes from the hyperfine Hamiltonian $\hat{\mathcal{H}}_{\rm HF} = a_{J} \, \hat{\mathbf{J}} \cdot \hat{\mathbf{I}}$, which, together with the Zeeman term $\hat{\mathcal{H}}_{B} = -	g_{J}	 \mu_{\mathrm{B}} \, \hat{\mathbf{J}} \cdot \mathbf{B}$, act as a perturbation to $\hat{\mathcal{H}}_{\rm CEF}$ in Eq.~\eqref{eq:cef_H_Stevens}.

	In our experiment we cannot directly probe the behaviour of the nuclear spins, and we cannot verify definitively the relevance of our findings with respect to the hyperfine aspects. If we were to strictly apply the criteria of Hammann and Manneville, then we would deduce that the molecular field is mainly of electronic nature and that the cooperative electro-nuclear polarization of the magnetic moment of the Tb$^{3+}$ ions will be relevant only at the lowest temperatures (our results in Tab.~\ref{tab:HammanQuantities} imply $x \gtrapprox 1.5$ in Fig. 3 of Ref.~[\onlinecite{hammann73}], which is to say that significant nuclear polarisation is only expected at $T\sim{}83$ mK or $T/T_N\sim{}0.33$)). However, one ought to conduct a dedicated study of instances of hyperfine driven quantum criticality at a level which is beyond the scope of this manuscript.

	The broadening and structure of the first excitation appearing at the lowest temperatures (Fig.~\ref{low_CEF_IRIS}) suggests that the crystal field level is developing into dispersive excitons, as expected in a singlet-singlet system\cite{leask94}, and it would be interesting to examine the extent of any mode-softening at the transition~\cite{zheludev96}.

	The crystal field states are of importance in the the acoustic Faraday effect (AFE)~\cite{thalmeier09,sytcheva10} and acoustic Cotton-Mouton effect~\cite{Low2014}, and in quantitative explanations of the elastic constants~\cite{Low2014} and may also be of importance in the thermal Hall effect. A detailed explanation of the acoustic Faraday effect exists~\cite{thalmeier09}, and depends in part on the character of the crystal field states of the main terbium site and their coupling with phonons. It is noted in Ref.~[\onlinecite{thalmeier09}] that the crystal field states were not precisely known, and a simplified scheme of cubic Tb$^{3+}$ site symmetry was adopted using a $\Gamma_{3}$ doublet and $\Gamma_{5}$ triplet, with specific field dependent components and matrix elements amongst them. A quantitative explanation of the temperature and field dependence of the elastic constants~\cite{Low13,Low2014} has been attempted using the parameters of Ref..~[\onlinecite{guillot85}], but, as described above, a more accurate description should be expected using the parameters presented here, particularly since the intersections and crossings of some levels are modified by the use of our parameters.

\section{Conclusions}\label{conclusions}

	The low-temperature magnetic structure of TGG is of the multiaxial antiferromagnet type. As closer examination has shown, the large anisotropy of the Tb$^{3+}$ magnetic moments and its direction can be explained by the time-reversal properties of the crystal-field states of this non-Kramers ion. The singlet ground-state of the Tb$^{3+}$ ion in TGG indicates the presence of perturbations allowing this induced-moment type of system to develop long-range order. Determination of the nature of the ordering transition still requires additional experimental investigation.

	Two striking anomalies can be recognised in the crystal-field spectra measured by neutron scattering. These are: substantial differences between observed and expected linewidths of low-lying excitations and the pronounced temperature-dependent structure of the quasidoublet ground-state (see Fig.~\ref{M_H}(b)). To address such effects, elucidating superexchange mechanisms between Tb$^{3+}$ ions, as well as coupling between magnetic and lattice degrees of freedom should be a priority for further studies. These investigations indeed introduce a \emph{collective} character by perturbing the microscopic states from those derived within a (semi-classical) single-ion approach. 

	In addition, the results of bulk measurements we have shown departures from a picture of non-interacting spins when cooling below 50~K. This is consistent with a suspected onset of the effects of interactions that may lead to a spin liquid phase of so far unknown character. The presence of a disordered, but correlated regime in TGG is not surprising given its moderate frustration ($f\sim{}32$). Nonetheless its appearance would be expected more in the range of temperatures $|\theta_{CW}|>T>T_{\rm N}$.

	Further investigations with single crystal samples that could provide the detailed $|\vec{Q}|$-dependence of the observed features are needed to determine precisely the perturbation affecting single-ion anisotropy and the exact character of the intermediate short-range order below 50~K in TGG. We believe that these would furthermore allow us to establish a consistent microscopic theory for the thermal Hall behaviour. By enriching the picture of the low temperature physics in TGG, the results we present here constitute the basis for a full microscopic description of the observed effect.

\begin{acknowledgments}
	The authors would like to express their gratitude to Victoria Garc\'{i}a Sakai for her assistance during IRIS experiment, and to Olivier C\'{e}pas, Michel Gingras, Michel Kenzelmann, Michiyasu Mori, and Mike Zhitomirsky for stimulating discussions and insightful suggestions. Neutron scattering experiments were carried out at the spallation neutron source ISIS at the Rutherford Appleton Laboratory, UK. The presented graphical representations of the crystallographic structure were prepared with \texttt{VESTA} software~\cite{vesta}.
\end{acknowledgments}

\bibliographystyle{apsrev}
\bibliography{TGG2019_arXiv_bibliography_2}

\begin{thebibliography}{63}
\bibitem[{kpo()}]{kpoint}
\bibinfo{note}{Here, $\mathbf{k}=(110)$ is identical to $\Gamma=(000)$ for
  BCC-type lattice.}
\end{thebibliography}

\appendix

\begin{widetext}

\section{System of coordinates} \label{sec:coordinates}
	In the rare earth garnets, each magnetic ion sits on one of the vertices joining the triangles of the two hyperkagome sublattices~\cite{ramirez01}, as represented in Fig.~\ref{tgg_struc}. From a \emph{local} point of view, the orthorhombic point-symmetry characterises the symmetry of the coordination environment or crystal field of a given $R^{3+}$ ion (not shown in Fig.~\ref{tgg_struc} -- details in Sec.~\ref{sec:cef}). From a \emph{global} point of view (i.e. with respect to the crystal-axes), the orientation of this local coordination environment varies from one site to the other. Throughout the whole garnet-lattice, according to the space-group $Ia\bar{3}d$, there are six orientations of the local axes with respect to the global cubic axes. In this work the local coordinates $\mathbf{x}_{i}, \mathbf{y}_{i}, \mathbf{z}_{i}$ are chosen to be parallel to the 2-fold rotation axes of the $D_2$ point group, so that with respect to the global $\mathbf{X,Y, Z}$ coordinates, they are:

\begin{subequations}
\begin{align}
\mathbf{x}_{1}	 &= [0, 0, 1], & \mathbf{y}_{1}&= \frac{1}{\sqrt{2}}[1, \bar{1},0],& \mathbf{z}_{1}	&=	\frac{1}{\sqrt{2}}[1,1,0];	\label{eq:LocCoord1}	\\	
\mathbf{x}_{2}	 &= [0, 1, 0], & \mathbf{y}_{2}&= \frac{1}{\sqrt{2}}[\bar{1}, 0,1],& \mathbf{z}_{2}	&=	\frac{1}{\sqrt{2}}[1,0,1];	\label{eq:LocCoord2}	\\
\mathbf{x}_{3}	 &= [1, 0, 0], & \mathbf{y}_{3}&= \frac{1}{\sqrt{2}}[0,1, \bar{1}],& \mathbf{z}_{3}	&=	\frac{1}{\sqrt{2}}[0,1,1];	\label{eq:LocCoord3}	\\
\mathbf{x}_{4}	 &= [0, 0, \bar{1}], &\mathbf{y}_{4}&= \frac{1}{\sqrt{2}}[\bar{1},\bar{1},0],& \mathbf{z}_{4}	&=	\frac{1}{\sqrt{2}}[\bar{1}, 1,0];	
\label{eq:LocCoord4}	\\
\mathbf{x}_{5}	 &= [0, \bar{1}, 0], &\mathbf{y}_{5}&= \frac{1}{\sqrt{2}}[\bar{1},0,\bar{1}],& \mathbf{z}_{5}	&=	\frac{1}{\sqrt{2}}[1, 0,\bar{1}];	
\label{eq:LocCoord5}	\\
\mathbf{x}_{6}	 &= [\bar{1}, 0, 0], &\mathbf{y}_{6}&= \frac{1}{\sqrt{2}}[0,\bar{1},\bar{1}],& \mathbf{z}_{6}	&=	\frac{1}{\sqrt{2}}[0,\bar{1},1].	
\label{eq:LocCoord6}		
 \end{align}
\label{eq:LocCoord}
\end{subequations}

This parametrisation allows for a coherent description of both single-ion physics and long range magnetic order. In the context of our study, each local coordinate system has $\mathbf{z}_{i}$ as its local quantisation axis and $\mathbf{x}_{i}$ coinciding parallel to the direction of the magnetic moment in the ordered phase.
The rotation from each system to the other can be obtained using the rotations 
$\mathbf{x}_{i}	=	\mathsf{R}_{i}	 \mathbf{X},
\mathbf{y}_{i}	=	\mathsf{R}_{i}	 \mathbf{Y},
\mathbf{z}_{i}	=	\mathsf{R}_{i}	 \mathbf{Z}$
, where $\mathsf{R}_{i}= (\mathbf{x}_{i}^{\mathsf{T}}, \mathbf{y}_{i}^{\mathsf{T}}, \mathbf{z}_{i}^{\mathsf{T}})$ is the rotation matrix and $ \mathbf{v}^{\mathsf{T}}$ represents the transpose of the vector $\mathbf{v}=[v_{X}, v_{Y}, v_{Z}]$.


\section{Eignevector structure of crystal-field states.} \label{eigenvectors_sec}
\begin{table*}
\centering
\setlength{\tabcolsep}{3.5pt} 
\def\arraystretch{2}
\caption{The 13 crystal-field eigenstates, $\ket{\psi_{i}}=\sum{c_{i,M_{J}} \ket{J, M_J}}$, expressed as expansions of the $\ket{M_J} \equiv \ket{J, M_J}$ eigenstates of $\hat{J}_{z}$, in the Russell-Saunders coupling scheme ($J=6, L=3, S=3$) for the ${^7F_6}$ ground multiplet of Tb$^{3+}$ ion. A blank entry means a zero coefficient.}
\label{eigenvectors}
\begin{tabular}[c]{c c ||c|| c|c|c|c|c|c|c|c|c|c|c|c|c|}
& & $E$ [meV] & $|-6\rangle$ & $|-5\rangle$ & $|-4\rangle$ & $|-3\rangle$ & $|-2\rangle$ & $|-1\rangle$ & $|0\rangle$ & $|1\rangle$ & $|2\rangle$ & $|3\rangle$ & $|4\rangle$ & $|5\rangle$ & $|6\rangle$\\ \hline
$|\psi_{0}\rangle$	& $\Gamma_3$ 	&	0.0 	& 		&-0.091 & 		&-0.026 & 		& 0.701 & 		& 0.701 & 		&-0.026 & 		&-0.091	& \\ \hline
$|\psi_{1}\rangle$	& $\Gamma_1$ 	&	0.22 	& 0.182 & 		& 0.122 & 		&-0.325 & 		&-0.832 & 		&-0.325 & 		& 0.122 & 		& 0.182 \\ \hline
$|\psi_{2}\rangle$	& $\Gamma_4$ 	&	4.50	&-0.594 & 		& 0.096 & 		& 0.371 & 		& 		& 		&-0.371 & 		&-0.096 & 		& 0.594 \\ \hline
$|\psi_{3}\rangle$	& $\Gamma_1$ 	&	5.25	& 0.683 & 		& 0.061 & 		&-0.085 & 		&-0.215 & 		&-0.085 & 		& 0.061 & 		&-0.683 \\ \hline
$|\psi_{4}\rangle$	& $\Gamma_4$ 	&	6.01	& 0.383 & 		& 0.115 & 		& 0.583 & 		& 		& 		&-0.583 & 		&-0.115 & 		&-0.383 \\ \hline
$|\psi_{5}\rangle$	& $\Gamma_2$ 	&	6.54	&	 	&-0.005 & 		& 0.502 & 		& 0.498 & 		&-0.498 & 		&-0.502 & 		& 0.005 & \\ \hline
$|\psi_{6}\rangle$	& $\Gamma_3$ 	&	26.40 	&	 	& 0.032 & 		& 0.706 & 		& 0.030 & 		& 0.030 & 		& 0.706 & 		& 0.032 & \\ \hline
$|\psi_{7}\rangle$	& $\Gamma_1$ 	&	28.27 	& 0.016 & 		& 0.415 & 		& 0.535 & 		&-0.290 & 		& 0.535 & 		& 0.415 & 		& 0.016 \\ \hline
$|\psi_{8}\rangle$	& $\Gamma_2$ 	&	34.26 	& 		& 0.566 & 		& 0.302 & 		&-0.298 & 		& 0.298 & 		&-0.302 & 		&-0.566 & \\ \hline
$|\psi_{9}\rangle$	& $\Gamma_3$ 	&	34.73 	&-0.019 & 		&-0.691 & 		& 0.148 & 		&		& 		&-0.148 & 		& 0.691 & 		& 0.019 \\ \hline
$|\psi_{10}\rangle$	& $\Gamma_4$ 	&	35.63 	& 		&-0.701 & 		& 0.035 & 		&-0.090 & 		&-0.090 & 		& 0.035 & 		&-0.701 & \\ \hline 
$|\psi_{11}\rangle$	& $\Gamma_2$ 	&	36.38 	& 		& 0.424 & 		&-0.397 & 		& 0.403 & 		&-0.403 & 		& 0.397 & 		&-0.424 & \\ \hline
$|\psi_{12}\rangle$	& $\Gamma_1$	&	38.20 	& 0.023 & 		& 0.556 & 		&-0.318 & 		& 0.422 & 		&-0.318 & 		& 0.556 & 		& 0.023 \\
\end{tabular}
\end{table*}

\end{widetext}

\end{document}